\documentclass[pdflatex,sn-mathphys-num]{sn-jnl}

\usepackage{graphicx}%
\usepackage{multirow}%
\usepackage{amsmath,amssymb,amsfonts}%
\usepackage{amsthm}%
\usepackage{mathrsfs}%
\usepackage[title]{appendix}%
\usepackage{xcolor}%
\usepackage{textcomp}%
\usepackage{manyfoot}%
\usepackage{booktabs}%
\usepackage{algorithm}%
\usepackage{algorithmicx}%
\usepackage{algpseudocode}%
\usepackage{listings}%

\usepackage{xcolor}

\begin{document}

\title[Article Title]{An Optimization Framework for Wide-Field Small Aperture Telescope Arrays Used in Sky Surveys}


\author[1]{\fnm{Wennan} \sur{Xiang}}\email{2024310284@link.tyut.edu.cn}

\author*[1]{\fnm{Peng} \sur{Jia}}\email{robinmartin20@gmail.com}
\equalcont{These authors contributed equally to this work.}

\author[2]{\fnm{Zhengyang} \sur{Li}}\email{zyli@niaot.ac.cn}

\author[3]{\fnm{Jifeng} \sur{Liu}}\email{jfliu@nao.ac.cn}

\author[1]{\fnm{Zhenyu} \sur{Ying}}\email{yingzhenyu3134@link.tyut.edu.cn}

\author[1]{\fnm{Zeyu} \sur{Bai}}\email{baizeyu1112@link.tyut.edu.cn}

\affil*[1]{\orgdiv{College of Physics and Optoelectronics}, \orgname{Taiyuan University of Technology}, \orgaddress{\street{No. 79 Yingze West Road}, \city{Taiyuan}, \postcode{030024}, \state{Shanxi}, \country{China}}}

\affil[2]{\orgdiv{Nanjing Institute of Astronomical Optics}, \orgname{Technology, Chinese Academy of Sciences}, \orgaddress{\street{No. 88 Hongshan Road}, \city{Nanjing}, \postcode{210094}, \state{Jiangsu}, \country{China}}}

\affil[3]{\orgdiv{National Astronomical Observatories}, \orgname{Chinese Academy of Sciences}, \orgaddress{\street{No. 20A Datun Road, Chaoyang District}, \city{Beijing}, \postcode{100012}, \state{Beijing}, \country{China}}}


\abstract{For time-domain astronomy, it is crucial to frequently image celestial objects at specific depths within a predetermined cadence. To fulfill these scientific demands, scientists globally have started or planned the development of non-interferometric telescope arrays in recent years. Due to the numerous parameters involved in configuring these arrays, there is a need for an automated optimization framework that selects parameter sets to satisfy scientific needs while minimizing costs. In this paper, we introduce such a framework, which integrates optical design software, an exposure time calculator, and an optimization algorithm, to balance the observation capabilities and the cost of optical telescope arrays. Neural networks are utilized to speed up results retrieval of the system with different configurations. We use the SiTian project as a case study to demonstrate the framework's effectiveness, showing that this approach can aid scientists in selecting optimal parameter sets. The code for this framework is published in the China Virtual Observatory PaperData Repository, enabling users to optimize parameters for various non-interferometric telescope array projects.}




\maketitle

\section{Introduction}\label{sec1}

\label{sect:intro}  
Time domain astronomy, a rapidly evolving field of research, focuses on the study of transient celestial phenomena. To effectively observe and analyze celestial objects such as electromagnetic counterparts of gravitational wave events, exoplanet systems, and supernovae, scientists require observation equipment with high temporal resolution and moderate detection sensitivity. To continuously capture images of astronomical objects within the required time resolution, researchers have proposed various telescope array configurations, including customized system projects such as CSTAR (Chinese Small Telescope ARray), FTN (the Falcon Telescope Network), and the SiTian project (known as Global Open Transient Telescope Array -- GOTTA now) \citep{article3, burd2005pi, article, cui2008antarctic, brown2013cumbres, Chun2018, tonry2018atlas, lokhorst2020wide, liu2021sitian}, as well as joint observational projects targeting existing telescope systems, such as GROWTH (Global Relay of Observatories Watching Transients Happen), GRANDMA (Global Rapid Advanced Network Devoted to the Multi-messenger Addicts), and ENGRAVE (Electromagnetic counterparts of gravitational wave sources at the Very Large Telescope) \citep{kasliwal2017global, antier2020multi, levan2020engrave}. These arrays normally consist of telescopes located at different locations with various longitudes or observing different regions of the sky from the same site, thereby enhancing the cadence and depth of observations. Within a telescope array, scientists employ diverse control strategies to maximize scientific output, optimizing the overall performance of telescope arrays and efficiency in data collection \citep{lampoudi2015integer, solar2016scheduling, bellm2019zwicky, jia2023observation}.\\

\begin{figure} [ht]
   \begin{center}
   \includegraphics[height=9cm]{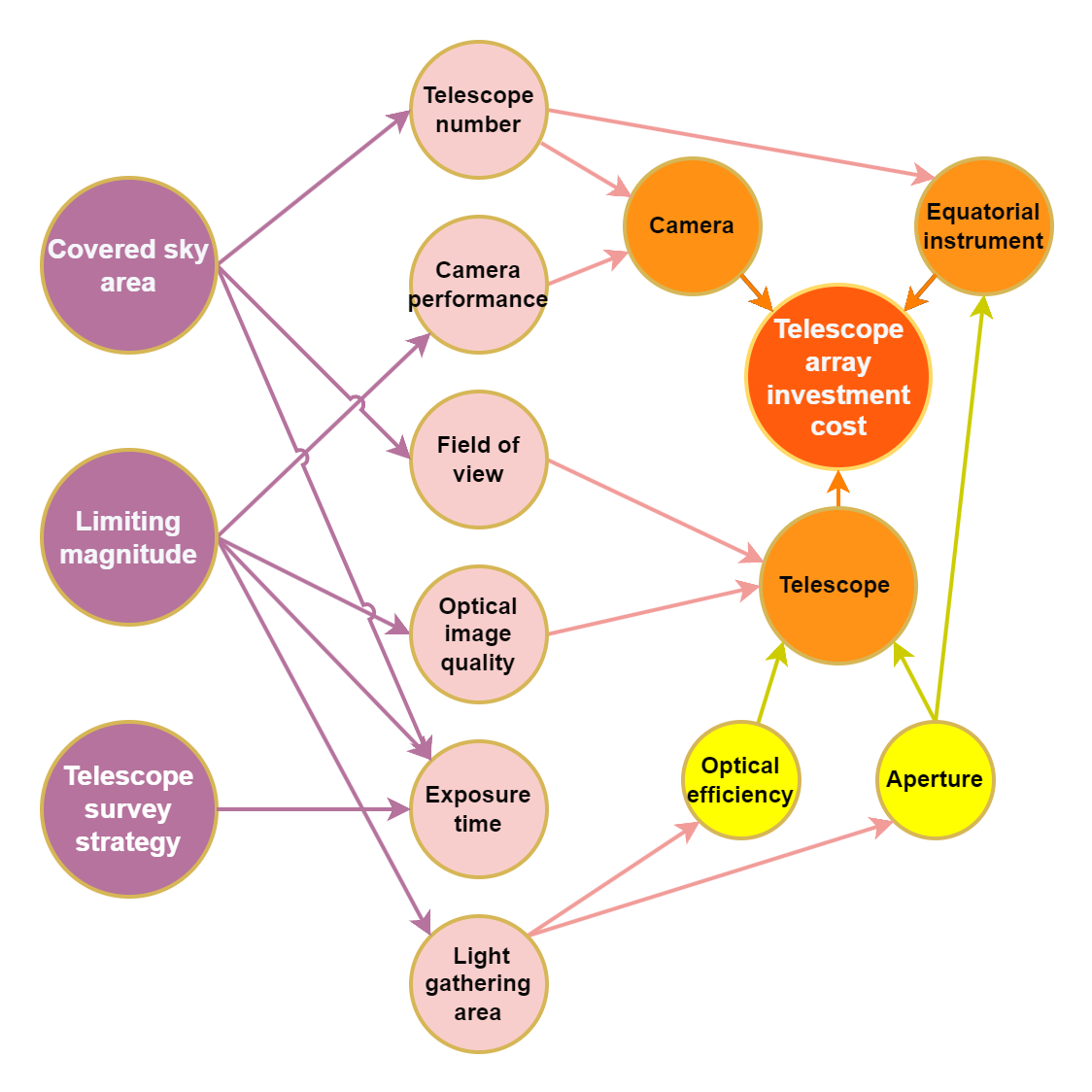}
   \end{center}
   \caption{This figure demonstrates the interconnections among various parameters of the optical system of the telescope, camera, and the equatorial mount in the telescope array, along with the overall performance and total cost of the system. Components with the same color stand for the same type. Components arranged from left to right, indicate different demand levels: the scientific requirements; the hardware parameters necessary to satisfy these demands; and the parameters enclosed within the yellow circle, which highlight the influence of light-gathering power on aperture size and obscuration ratio. These parameters show the specifications for the three primary hardware components in a telescope array—telescopes, detectors, and equatorial mounts—resulting in an estimated total cost for the optical system. The figure visually illustrates how different design elements and specifications of individual telescopes enhance the overall capability of the array in observing transient celestial events.}
   {\label{fig:1}} 
   \end{figure}

\begin{figure} [ht]
   \begin{center}
   \includegraphics[height=7.5cm]{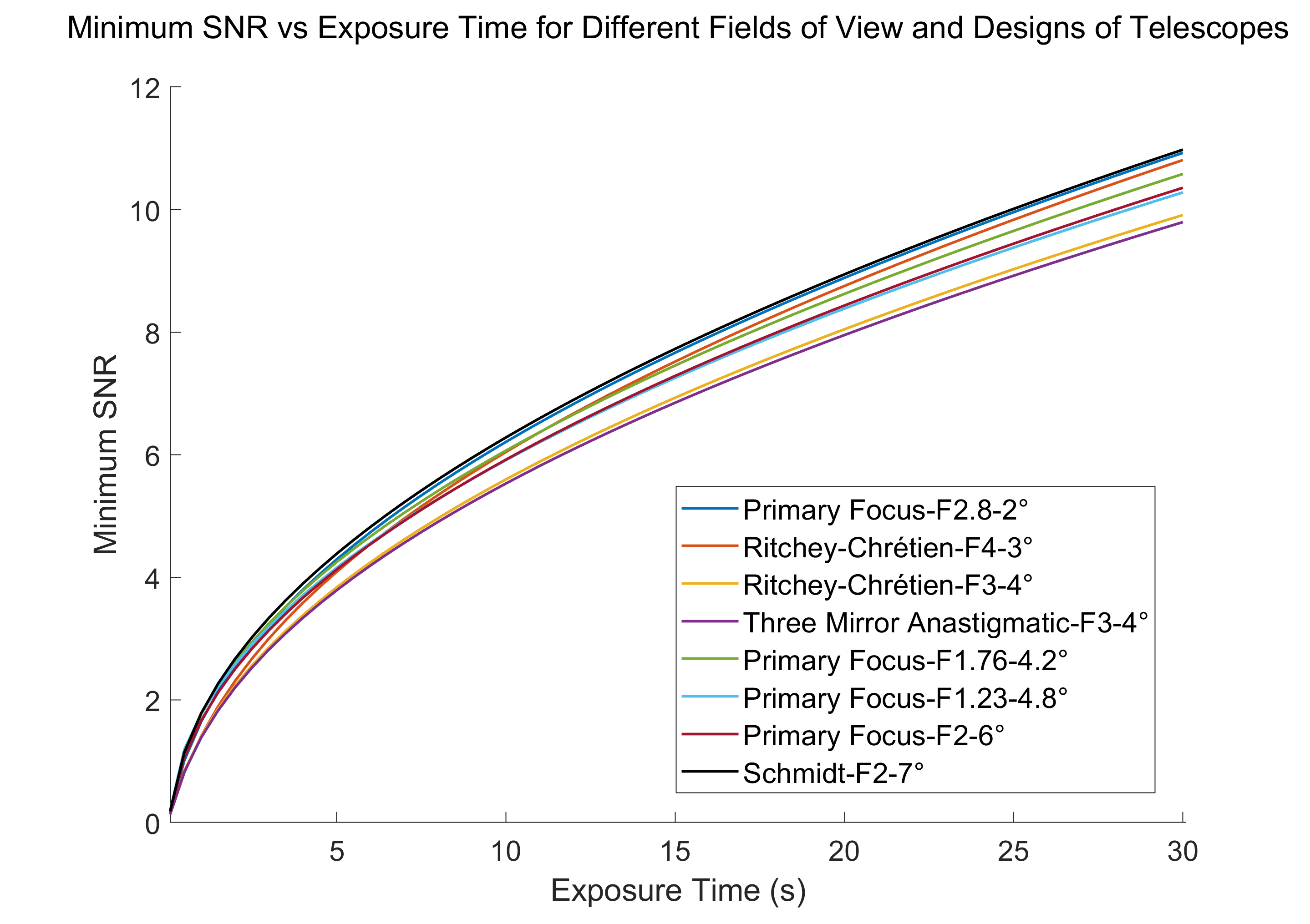}
   \end{center}
   \caption{Variation trend of the minimum signal-to-noise ratio (SNR) of simulated images from 1-meter diameter telescopes with different designs under varying exposure times.}
   {\label{fig:relationshape_para}} 
   \end{figure}

Designing a optical telescope array that contains several non-interferometric wide-field telescopes for sky survey projects involves carefully selecting numerous free parameters. Determining optimal parameters for such an array, which can satisfy scientific requirements while minimizing costs, presents a significant challenge. Various factors have complex interrelationships that affect overall performance. Figure \ref{fig:1} illustrates the relationships between cost and parameters of an optical telescope array. To provide a more intuitive understanding of the relationships among these parameters, we have plotted Figure \ref{fig:relationshape_para}, which shows the relative variation trend of the minimum signal-to-noise ratio (SNR) of telescopes with different fields of view and designs under varying exposure times used in this study. The figure presents the minimum SNR variation for 1-meter diameter telescopes at different exposure times within the designed field of view (with the observational environment set to the default parameters established in the subsequent sections of this paper, the target star magnitude set to 21, and the sky background magnitude set to 22.3). Through this illustration, we can more clearly see the relationships between different design choices, exposure times, and imaging quality. There are also more interconnections among other parameters. For example, telescopes with an optical design enabling a larger field of view would require fewer units to meet scientific needs. However, such wide-field-of-view telescopes tend to be more expensive. Consequently, developing a framework that facilitates the optimization of the telescope array parameters becomes essential to effectively address these challenges \citep{xiang2022automatic}.\\

In this paper, we present a framework for automatically exploring the parameter space of telescope arrays and obtaining optimal parameter sets based on given telescope designs and scientific requirements. The framework comprises three key components: an exposure time calculator, an optical system simulator (or optical design software), and an optimizer that searches for optimal telescope array parameters. Optical simulations are conducted using ZEMAX, a widely recognized optical design software renowned for its accurate ray-tracing capabilities in simulating optical systems. In conjunction with ZEMAX, MATLAB is utilized to connect with ZEMAX, perform exposure simulations. Given that parameter fine-tuning and repeated calculations can be time-consuming, we also propose the use of a Back Propagation Neural Network (BP neural network) to obtain the performance of the telescope under various observational conditions. Section \ref{sec2} provides a detailed discussion of the framework design, while Section \ref{sec3} demonstrates the prototype design for the SiTian project with the framework \citep{liu2021sitian}. Lastly, Section \ref{sec4} summarizes our work and outlines future research directions.\\

\section{Methods}\label{sec2}

\subsection{Overall Structure of the Framework}
\label{sec21} 
\begin{figure*}[htbp]
    \centering
    \begin{minipage}[t]{0.24\textwidth}
        \centering
        \includegraphics[width=\textwidth]{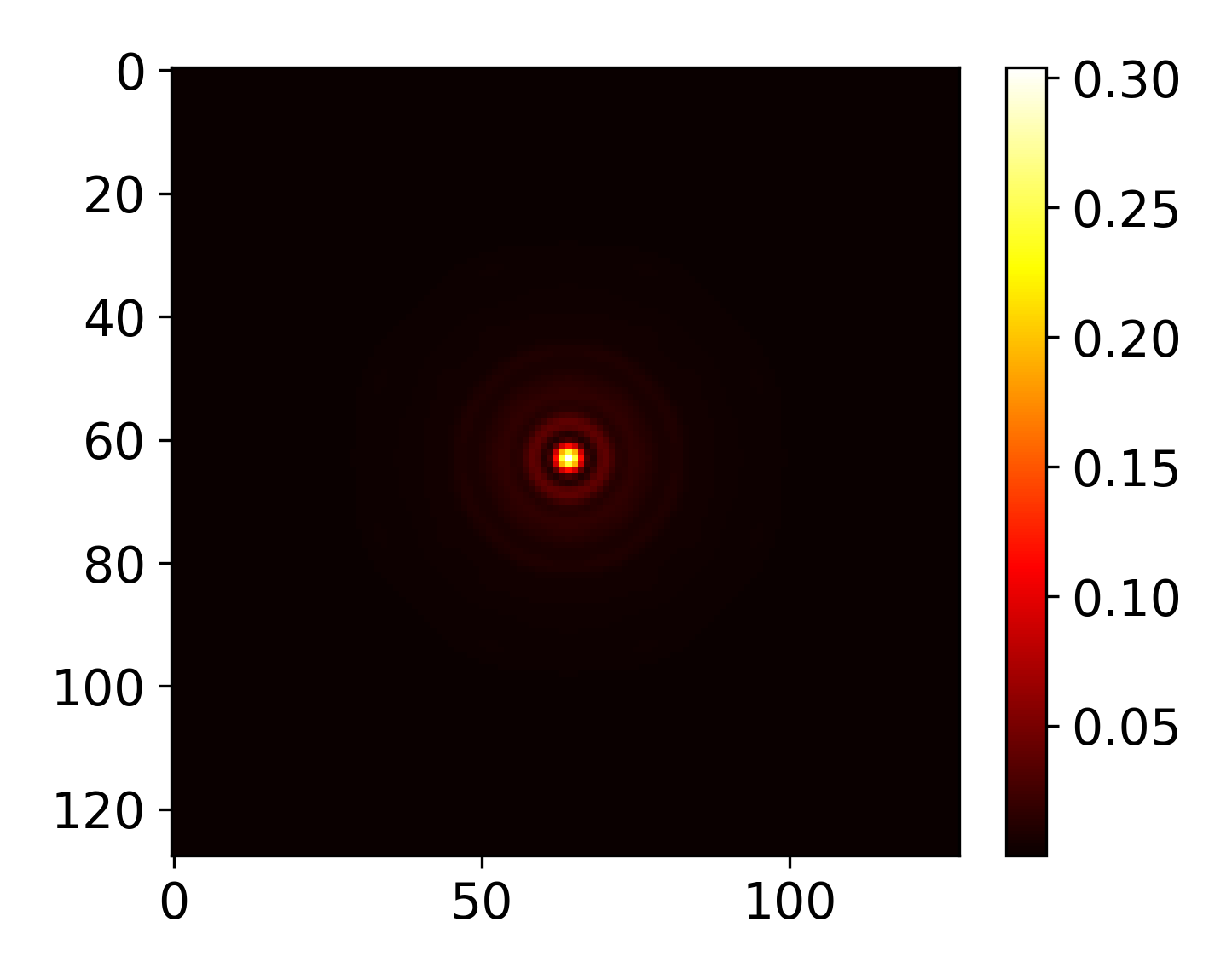}
        \textbf{(a)}
    \end{minipage}
    \hfill
    \begin{minipage}[t]{0.24\textwidth}
        \centering
        \includegraphics[width=\textwidth]{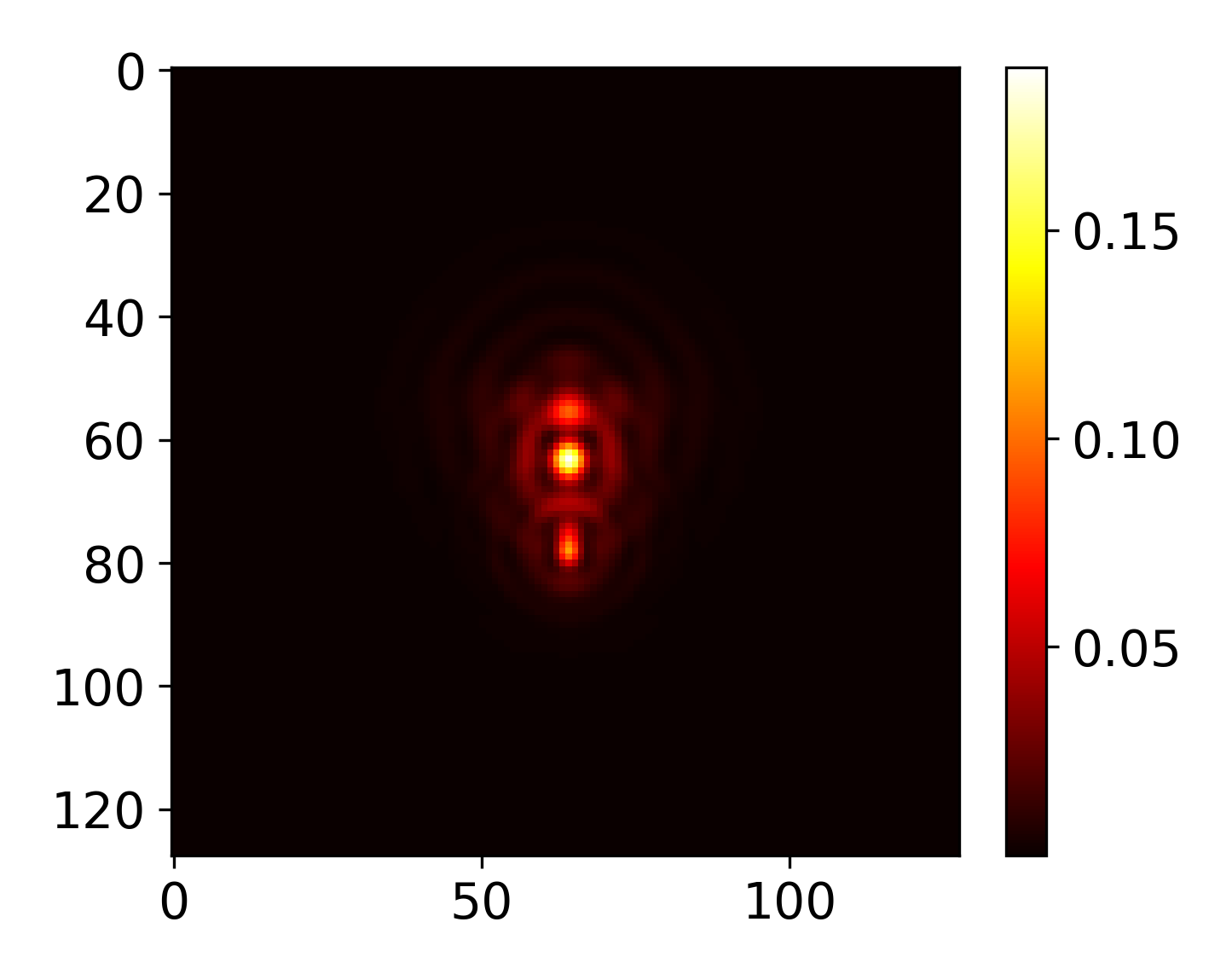}
        \textbf{(b)}
    \end{minipage}
    \hfill
    \begin{minipage}[t]{0.24\textwidth}
        \centering
        \includegraphics[width=\textwidth]{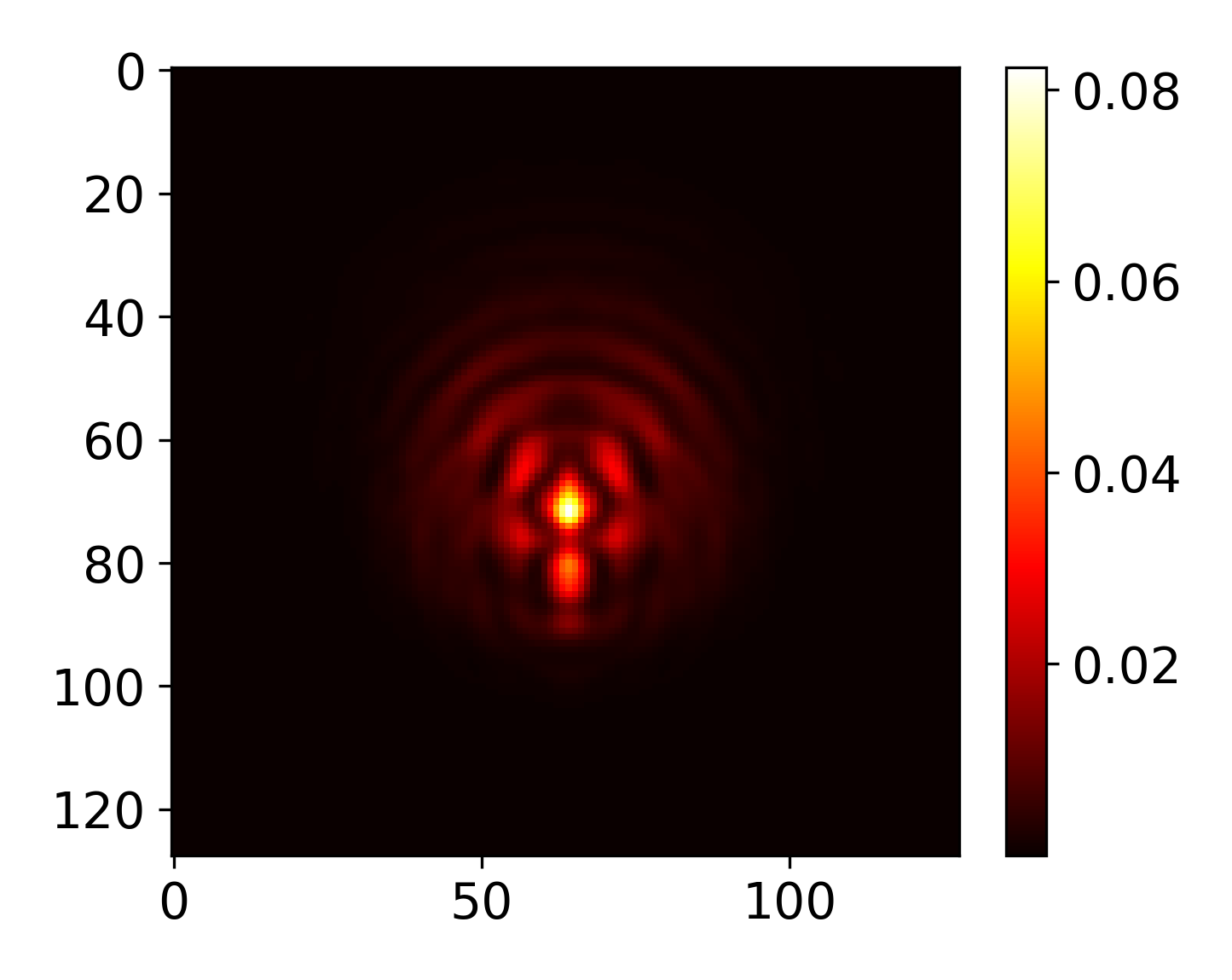}
        \textbf{(c)}
    \end{minipage}
    \hfill
    \begin{minipage}[t]{0.24\textwidth}
        \centering
        \includegraphics[width=\textwidth]{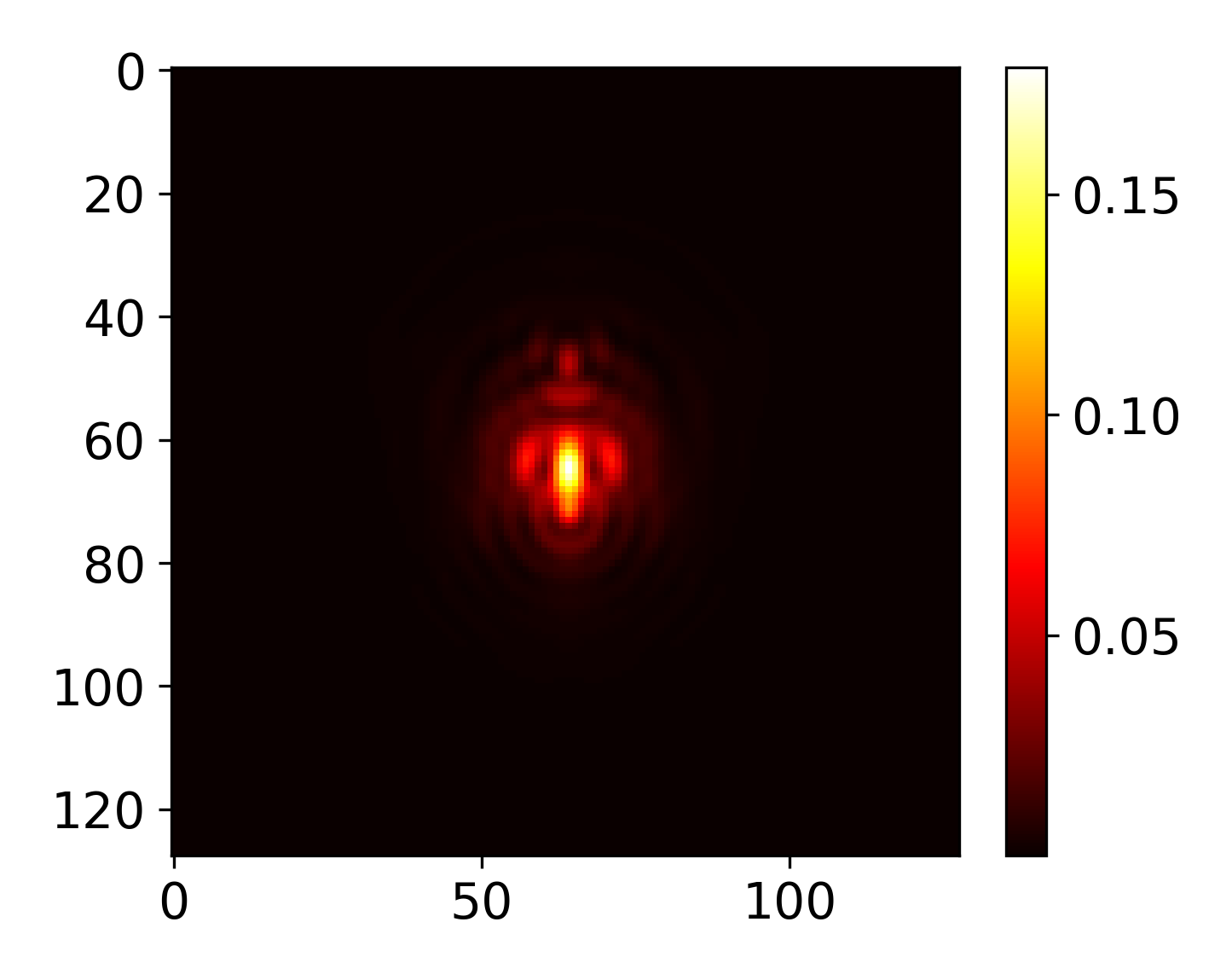}
        \textbf{(d)}
    \end{minipage}
    \caption{\label{fig:wide_field_Tel_PSF} PSF variation from the optical axis to the maximum field of view of a telescope with design (Primary Foucus-F2-$6^{\circ}$).}
\end{figure*}

In this subsection, we provide a concise overview of the optical system simulator, the exposure time calculator, the system optimizer, and their interrelations within the framework. The optical system simulator generates Point Spread Functions (PSFs) across different fields of view based on the specified optical design, as shown in Figure \ref{fig:wide_field_Tel_PSF}. Given that telescopes used in time-domain astronomy observations typically have a wide field of view, their PSFs vary across different field of views  \citep{zhao2012lamost, bellm2018zwicky, ivezic2019lsst}. To evaluate the performance of a telescope array accurately, we divide the field of view into smaller segments and generate PSFs for each section. These PSFs are then convolved with PSFs induced by atmospheric turbulence, resulting in over-sampled optical PSFs. Subsequently, these optical PSFs are forwarded to the exposure time calculator component. This component considers various factors, including the observation waveband, different types of noise from the sky background and the detector, pixel scale, and atmospheric conditions, to provide a comprehensive analysis of the performance of the telescope.\\

The exposure time calculator generates simulated observation images and calculates the signal-to-noise ratio (SNR) of a celestial object for a given exposure time. The performance of the telescope array is assessed using both the optical system simulator and the exposure time calculator. The system optimizer, responsible for exploring the parameter space, interacts with these two components. This system optimizer includes two algorithms: a brute-force search algorithm and a parameter retrieval algorithm based on BP neural network. By utilizing these two algorithms, it is possible to determine a set of parameters that both meet scientific requirements and minimize costs, thereby optimizing the overall efficiency and effectiveness of the telescope array and providing optical experts with assistance in parameter selection.\\

We have developed two versions of the framework: one for Windows and the other for Linux. The Windows version employs MATLAB as the primary computational tool to generate simulated images that correspond to the system, assess imaging quality, and estimate system costs. It interacts with ZEMAX through the ZOS-API (Zemax OpticStudio Application Programming Interface), which is supported by ZEMAX \citep{griffith2006talk, zemaxzos}. The Linux version relies on Python as the foundational computational platform and utilizes POPPY (Physical Optics Propagation in Python), a third-party library specifically designed for simulating and analyzing optical wave propagation \citep{perrin2016poppy}. However, when modifying the configurations of an optical system, it becomes necessary to verify the model built with POPPY, which requires frequent manual interventions. Therefore, to actively connect the optical system simulator with the optimization algorithm, further modifications to the Linux system are necessary. In this paper, we focus on performing our computations using the Windows version of the framework. We will provide detailed information about the optical system simulator, the exposure time calculator, and the system optimizer in the subsequent subsections.\\

\subsection{The Optical System Simulator}
\label{sec22}
The optical system simulator generates PSFs of telescopes based on their optical design. To account for the detector's pixel scale and the influence of atmospheric turbulence on PSFs, we generate oversampled PSFs in this step, ensuring an effective spatial sampling rate. Typically, the spatial sampling rate is set to be at least twice smaller than the pixel scale of the camera. In practical applications, we initially acquire all necessary optical structures of the telescope using MATLAB. Following this, we import the corresponding telescope into ZEMAX, where we adjust its optical design parameters and calculate its PSFs. Utilizing the PSFs and effective focal length obtained from ZEMAX, we then perform calculations and generate simulated images in MATLAB. To establish a dynamic data connection between MATLAB and ZEMAX, enabling seamless integration and efficient data exchange between the two simulation and computational tools.\\

After obtaining the PSF of the optical system under different fields of view, we simulate the effect of atmospheric conditions on imaging during long exposures using the Moffat model \citep{Trujillo2001}. The Moffat model is a mathematical model used to describe the intensity distribution of point sources in an optical system. This model is particularly useful in astronomical applications, enhancing the accuracy and reliability of the simulations. Subsequently, the optical PSF is convolved with the atmosphere-induced PSF to generate the final optical PSF. These final PSFs are then utilized in the exposure time calculator to generate simulated images and calculate their corresponding SNR. It is important to note that, we partition the field of view into discrete grids to equally represent different field of views \citep{Jia2018}. This approach allows for a more comprehensive assessment of the performance of the telescope, accounting for variations of image quality across different parts of the field.\\

\subsection{The Exposure Time Calculator}
\label{sec23}
\begin{figure} [ht]
   \begin{center}
   \includegraphics[height=8cm]{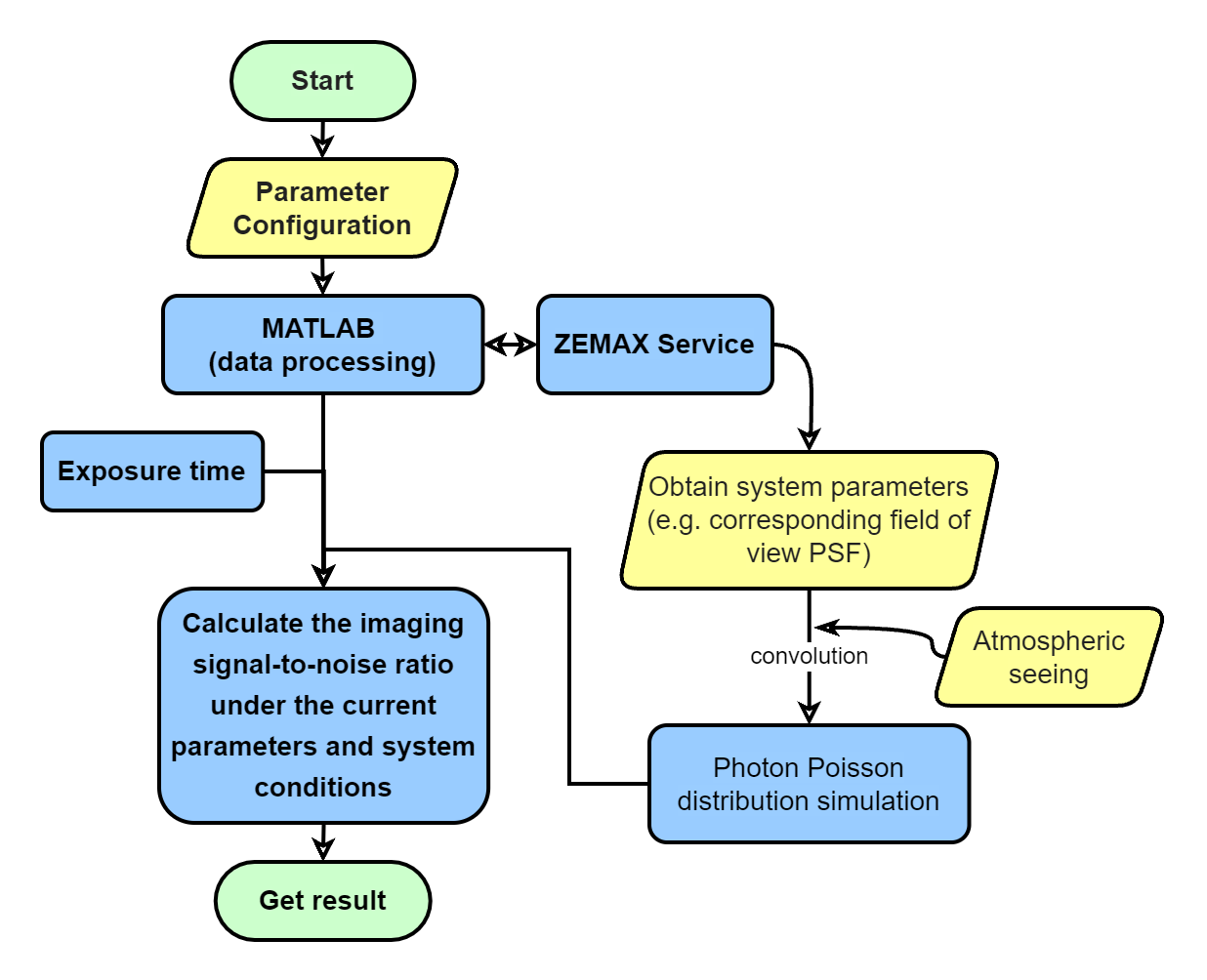}
   \end{center}
   \caption{The flowchart of the exposure time calculator.}
   {\label{fig:2}} 
   \end{figure}

The exposure time calculator is employed to generate final observation images based on the provided PSFs, observation conditions, and characteristics of celestial objects. It also calculates the SNR of the image for the corresponding exposure time. The flowchart of the exposure time calculator is depicted in Figure \ref{fig:2}. As illustrated, the exposure time calculator loads the observation parameters and utilizes MATLAB to preprocess these parameters. Specifically, the parameters associated with the optical design of telescopes are transmitted to the optical system simulator to generate the corresponding PSFs. Once the PSFs are obtained, we generate observation images and use the SNR to evaluate the image quality. Further details of this process will be discussed below.\\

\begin{figure*}[ht]
\begin{center}
\includegraphics[height=10cm]{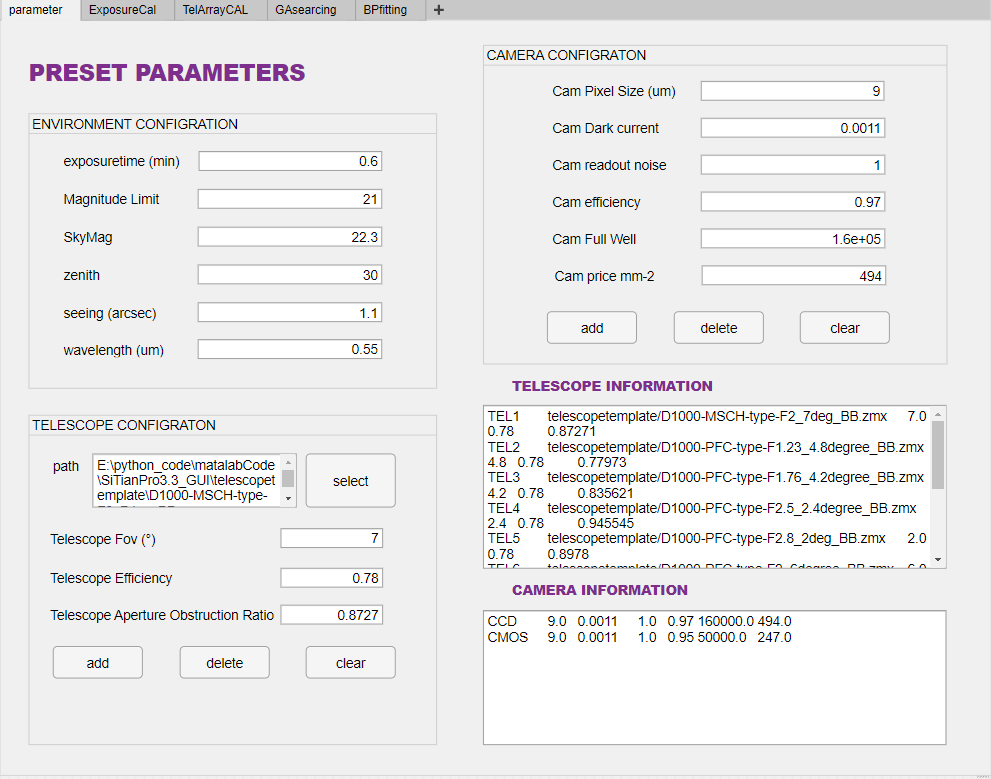}
\end{center}
\caption{User interface of the framework.}
{ \label{fig:3}} 
\end{figure*}

Some input parameters that are directly related to the observation environment with their default values are presented in Table~\ref{tab:1}. These parameters include the zenith angle, seeing conditions, observation waveband, camera type, telescope transmittance and the image overlap between different telescopes. The zenith angle has effect to the thickness of atmosphere along the observation direction. The seeing condition describes blurring effects to the final observation images. The observation waveband affects the number of photons received from celestial objects. Different types of cameras exhibit varying noise characteristics and quantum efficiencies. Users can set parameters corresponding to the camera performance. In this paper, we show a customized CCD camera as the detector in Table~\ref{tab:2}. Observation images can be generated based on these specific camera parameters. The price of the customized camera will vary with changes in the telescope optical system, which will be discussed later in the text. The telescope's transmittance is assumed as an input value, describing the telescope's efficiency. The image overlap refers to the number of shared pixels when stitching together multiple telescope fields of view. In addition to these parameters, there are many other parameters that users can modify or define with the user interface system, as shown in Figure~\ref{fig:3}.\\

\begin{table}[ht]
\caption{Some input parameters and their default values.} 
\label{tab:1}
\begin{tabular}{@{}llll@{}}
\toprule
System parameters & Value  \\
\midrule
Zenith $\ (\deg)$           & 30  \\
Seeing \ ('')               & 1.5   \\
Wave\ Band                  & V band \\
Camera                      & CCD   \\
Transmittance               & 0.78  \\
Image\ Overlap \ (pixel)    & 5     \\
\botrule
\end{tabular}
\end{table}

\begin{table}[ht]
\caption{Parameters of the CCD cameras defined in framework, which is obtained from vendors. However it should be noted that the parameters provided are for reference only and could be adjusted according to specific conditions in practical applications.} 
\label{tab:2}
\begin{tabular}{@{} p{5cm} p{1.5cm} @{}}
\toprule
Type of camera                           & CCD       \\  
\midrule
Pixel\ size\ ($\mu$m)                   & 9            \\ 
Quantum\ efficiency\ (V band)            & 0.97      \\ 
Dark\ current\ ($\mathrm{e}^{-} \mathrm{s}^{-1} \mathrm{pixel}^{-1}$) & 0.0011   \\ 
Readout\ noise\ ($\mathrm{e}^{-} \mathrm{pixel}^{-1}$)        & 1            \\ 
FullWell\ Depth\ ($\mathrm{e}^{-} \mathrm{pixel}^{-1}$)       & 160000   \\
Relative\ Unit\ Cost\ (cost\ $\mathrm{mm}^{-2}$)          & 494       \\ 
\botrule
\end{tabular}
\end{table}

Once the environment parameters are configured, we calculate the PSFs of the current optical system across different fields of view using the optical system simulator. To conduct comprehensive simulations, we generate observation images that consider various factors, including the photometric system, atmospheric extinction at the observation site, telescope aperture, and camera parameters. We utilize the Johnson-Morgan-Cousins UBVRI photometry system and the extinction coefficients of La Palma on a typical dust-free night \citep{bessell1990ubvri, buton2013atmospheric, garcia2010atmosphere}. Once the atmospheric extinction coefficients and observation band are determined, we can calculate the apparent magnitude of a star with known magnitude after its light passes through the atmosphere, as well as the apparent magnitude of the sky background, using Equation~\ref{equation1},
\begin{equation}
\label{equation1}
M = M_\mathrm{origin}+1/\text{cos}(\theta_\mathrm{zenith})\cdot C_\mathrm{extinction},
\end{equation}
where $M$ represents apparent magnitude obtained by the telescope, $M_\mathrm{origin}$ represents the original apparent magnitude, $\theta_\mathrm{zenith}$ represents the angular distance between the observed target and the zenith, affecting the path length of starlight through the atmosphere. The atmospheric path length is defined as 1 when $\theta_\mathrm{zenith}=0$. For cases where $\theta_\mathrm{zenith}$ is less than 60 degrees, atmospheric path variation can be approximated by $1/\text{cos}(\theta_\mathrm{zenith})$. The extinction coefficient $C_\mathrm{extinction}$ is a proportional constant that describes the impact of atmospheric extinction on apparent magnitude when the observed target is at the zenith. Its value depends on environmental conditions, time, the observational waveband, and the site condition.\\

Given the apparent magnitude and the photon density at zero magnitude, we can calculate the photon density of a celestial object or sky background. Based on the relation between photon densities and the reference photon number of a zero-magnitude star provided in the corresponding photometry system, we can calculate the number of photons corresponding to stars of different apparent magnitudes \citep{pogson1856magnitudes, henden1982astronomical}. According the telescope aperture, field of view, obstruction ratio, and telescope's transmittance. we can calculate the number of photons received by the telescope with Equation~\ref{equation3} and ~\ref{equation4},
\begin{equation}
\label{equation3}
N_\mathrm{source} = F(m_\mathrm{source}) \cdot t \cdot  \pi \cdot (\frac{D}{2} )^{2}\cdot R_\mathrm{aperture}\cdot T_\mathrm{telescope},
\end{equation}
\begin{equation}
\begin{aligned}
\label{equation4}
N_\mathrm{sky} = F(m_\mathrm{sky} \cdot \Omega_\mathrm{telescope}) \cdot t \cdot \pi \cdot (\frac{D}{2})^{2}\cdot R_\mathrm{aperture}\cdot T_\mathrm{telescope},
\end{aligned}
\end{equation}
where $N_\mathrm{source}$ and $N_\mathrm{sky}$ stand for the number of photons received from the celestial object and the sky background, respectively, by the detector, $N_\mathrm{sky}$ is uniformly distributed across the target surface. $t$ stands for the exposure time, $D$ stands for the diameter of the telescope, $R_\mathrm{aperture}$ stands for the effective aperture obstruction ratio (ratio of the aperture after we consider the obstruction), $T_\mathrm{telescope}$ stands for the transmittance of the telescope. $m_\mathrm{source}$ and $m_\mathrm{sky}$ denote the apparent magnitudes of the celestial object and the sky background, respectively. Because the photon flux of celestial objects is expressed in units of $\text{photons} \ \text{s}^{-1} \ \text{cm}^{-2}$, and the photon flux of the sky background light is the integral of the flux per unit solid angle over the entire field of view of the telescope, we could obtain the photon flux of stars $F(m_\mathrm{source})$ and backgrounds $F(m_\mathrm{sky} \cdot \Omega_\mathrm{telescope})$ with known magnitudes, where $\Omega_\mathrm{telescope}$ represents the solid angle of the telescope's field of view.\\

\begin{figure*}[htbp]
    \centering
    \begin{minipage}[t]{0.24\textwidth}
        \centering
        \includegraphics[width=\textwidth]{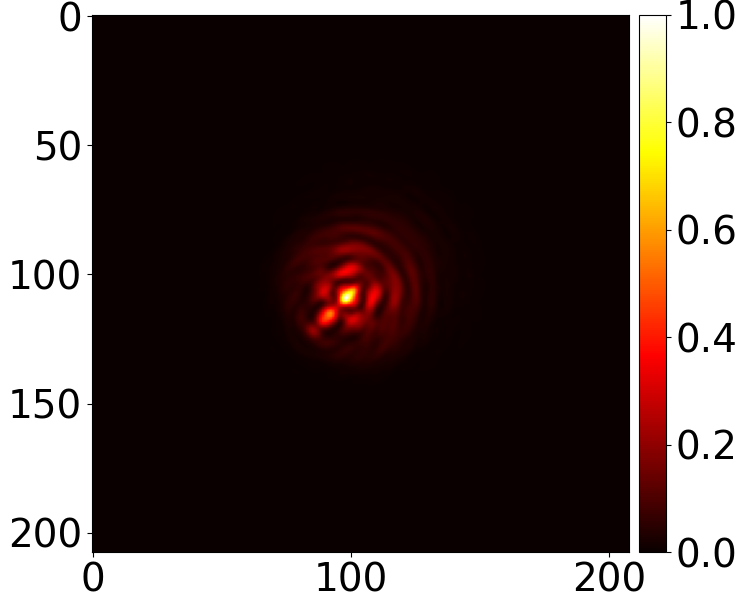}
        \textbf{(a)}
    \end{minipage}
    \hfill
    \begin{minipage}[t]{0.24\textwidth}
        \centering
        \includegraphics[width=\textwidth]{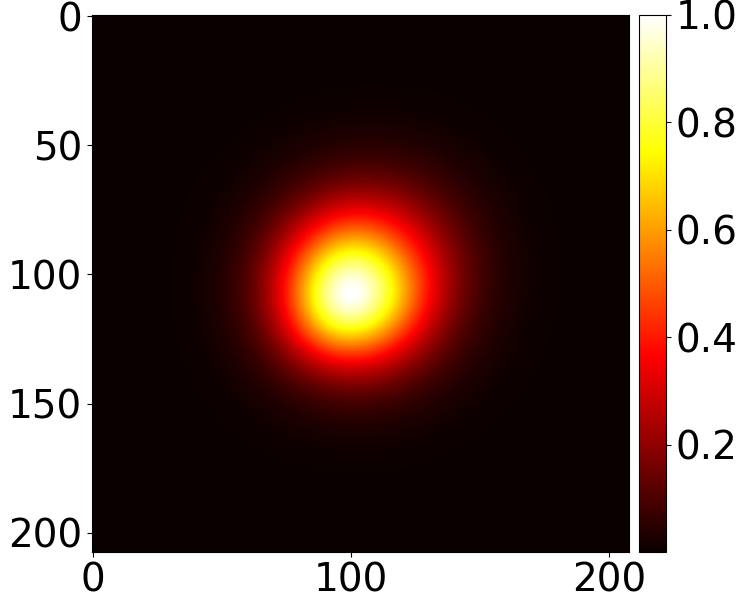}
        \textbf{(b)}
    \end{minipage}
    \hfill
    \begin{minipage}[t]{0.24\textwidth}
        \centering
        \includegraphics[width=\textwidth]{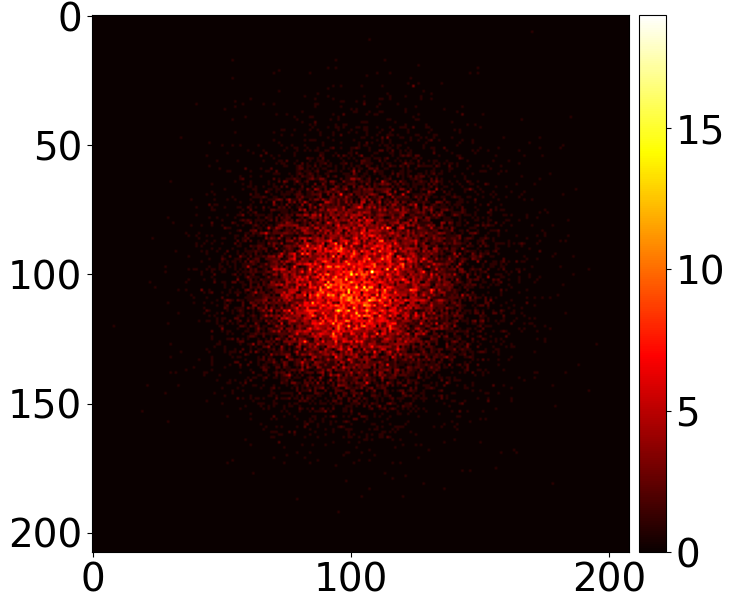}
        \textbf{(c)}
    \end{minipage}
    \hfill
    \begin{minipage}[t]{0.24\textwidth}
        \centering
        \includegraphics[width=\textwidth]{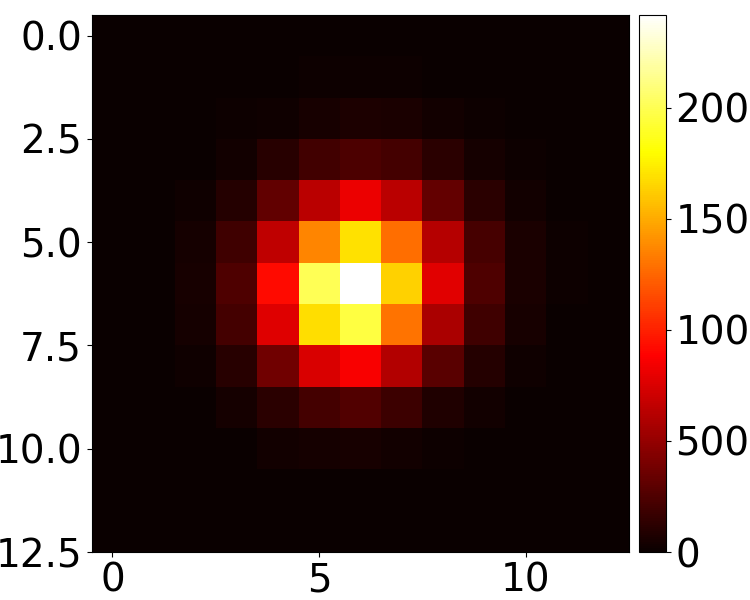}
        \textbf{(d)}
    \end{minipage}
    \caption{\label{fig:4} Images generated by the exposure time calculator during different steps. (a) displays the original PSF obtained from the optical system simulator. (b) shows the long exposure PSFs, which account for atmospheric effects over extended observation periods. (c) presents the photon image, demonstrating the distribution of detected photons based on the the PSF, the atmospheric effects and number of photons from the celestial object. Finally, (d) shows the simulated observation image, which incorporates all relevant factors including detector noises and spatial sampling rate.}
\end{figure*}

Using the calculated photon count, we employ the PSF as a probability distribution to generate photon images. We then incorporate sky background photons into these simulated images and simulate the effects of long exposure by convolving with Moffat Model. Next, we discretize the photon images to align with the pixel scale of the camera. We apply camera-specific parameters to produce the final simulated images, including photon-to-electron conversion, addition of dark current and readout noise, and spatial digitization. Following these steps, we obtain simulated images as illustrated in Figure~\ref{fig:4}.\\

In the exposure time calculator, we further calculate SNR of a celestial object to evaluate the performance of the telescope array, as shown in Equation \ref{equation5},
\begin{equation}
\begin{aligned}
\label{equation5}
\text{SNR} = \frac{N_\mathrm{source}}{\sqrt{N_\mathrm{source}+(N_\mathrm{sky,pixel\ size}+\sigma _\mathrm{instrument}^{2}) n_\mathrm{eff} }},
\end{aligned}
\end{equation}
where $N_\mathrm{source}$ is the total number of photons from the celestial object received by the detector, $N_\mathrm{sky,pixel\ size}$ is the number of photons contributed by the sky background per pixel on the detector, $\sigma _\mathrm{instrument}$ represents the variance of the instrumental noise per pixel, which includes dark current and readout noise, and $n_\mathrm{eff}$ is defined by the PSF and describes the number of effective pixels affected by a given celestial object during the imaging process, as shown in Equation \ref{equation6},
\begin{equation}
\begin{aligned}
\label{equation6}
n_\mathrm{eff}=4 \pi \cdot(\text{FWHM}/\text{PixelScale})^{2},
\end{aligned}
\end{equation}
where pixelScale refers to the camera's pixel scale. We fit the PSF with a Gaussian function following the method outlined by \citet{king1971profile}. In our algorithm, the default photometric aperture size is set to be four times the full width at half maximum (FWHM) of the PSF. This configuration ensures that nearly all photons from the target source's image is captured within the aperture, as suggested by \citet{TWJZ201204006}. We will check if the design of the telescope array satisfies the imaging quality requirements with the above equations.\\

\subsection{The System Optimizer}
\label{sec24}
The system optimizer employs two algorithms: a brute-force search algorithm, and a parameter retrieval algorithm based on BP neural networks. These two algorithms can effectively explore the parameter space for a telescope array. There are four principles that we would consider in the system optimizer:
\begin{itemize}
    \item Within the framework of this paper, the selection of telescope array parameters is optimized with cost as a key consideration. The goal is to ensure that the optical system of the telescope array can perform sky surveys to the specified depth within the designated observation period while minimizing costs.
    \item The minimum observation time for a cadence, encompassing both exposure and readout times, is dictated by scientific requirements. In this study, it is defined as the exposure time necessary to observe a star of a predetermined magnitude and achieve the desired SNR.
    \item The total observed sky area is determined by three factors: the field of view of individual telescopes, the number of telescopes in the array, and the number of observations conducted within a single cadence.
    \item The cost of each telescope in the array is calculated by considering three main components: the cost of the telescope, the cost of the camera and the cost of the annual maintenance. The cost of the camera, as outlined in Table~\ref{tab:2}, is associated with parameters including the telescope's aperture, effective focal length, and field of view, all of which affect the size of the detector. Using the pixel scale of the detector and the field of view, we can determine the number of pixels needed for the camera. Additionally, the dimensions of the detector can be calculated based on the focal length and the F-number. By combining the pixel count with the dimensions of the detector, we can derive the overall cost of the customized camera. The telescope cost, which is intricately linked to factors such as materials, structure, and whether they are produced on a large scale. The cost of the telescope's optical system increases exponentially with the aperture size, with an exponent ranging from 2 to 3 \citep{stahl2016multivariable, stahl2020parametric}. In the optical system of the telescope array discussed in this paper, considering mass production of small telescopes, we assume the growth exponent to be 2. The annual maintenance cost is estimated as a specific percentage of the total cost of all telescopes in the array.
\end{itemize}
With above principles, we could use algorithms in the optimizer to investigate the parameter space.\\

\subsubsection{Brute Force Search Algorithm}
\label{sec241}
The brute-force search algorithm is the initial approach implemented in our framework. This straightforward method is used to determine the parameters of a telescope array when we have knowledge of the prototype design of telescopes employed in the array. The entire process of the brute-force search involves several steps. Given the parameters of the optical design and camera, we obtain the PSF of the optical system. Using the exposure time calculator proposed in Section~\ref{sec23}, we then calculate the minimum exposure time needed for a telescope to meet the SNR requirements. Simultaneously, we determine the sky area that a telescope can observe within a given observation interval. By combining the exposure time and the requirement for the total sky survey area, we can calculate the total number of telescopes needed for the observations and subsequently determine the total cost of the telescope array. Through searching across all different aperture sizes of telescope designs and the total costs of the array under various camera parameter configurations, we can identify different configurations that meets the observational requirements.\\

Given the observation interval for an optical telescope array system, each telescope can perform multiple exposures and observe different sky regions within a single observation cycle $T_\mathrm{interval}$. We define the parameter $N_\mathrm{stack}$ to represent the number for an individual telescope points to distinct sky regions within one cycle. Additionally, we define $N_\mathrm{pointing}$ as the minimum number of pointing obtained by all telescopes within a single observation cycle to cover the sky regions necessary to meet scientific observation requirements. Based on the total sky solid angle to be observed $\Omega_\mathrm{all fov}$, the overlapping field of view due to multiple observation pointing deviations $\Omega_\mathrm{overlap}$, and the telescope’s field-of-view solid angle $\Omega_\mathrm{telescope}$,  $N_\mathrm{pointing}$ can be determined as $(\Omega_\mathrm{all fov} + \Omega_\mathrm{overlap}) /\ \Omega_\mathrm{telescope}$. We can obtain the number of telescopes $N_\mathrm{telescope}$ needed for the optical system of the telescope array using Equations~\ref{stack} and \ref{N_{telescope}}:
\begin{equation}
\begin{aligned}
\label{stack}
N_\mathrm{stack} = \frac{T_\mathrm{interval}}{T_\mathrm{exposure}+ T_\mathrm{slewing}},
\end{aligned}
\end{equation}
\begin{equation}
\begin{aligned}
\label{N_{telescope}}
N_\mathrm{telescope}=\text{ceil}(\frac{N_\mathrm{pointing}}{N_\mathrm{stack}}),
\end{aligned}
\end{equation}
where $T_\mathrm{exposure}$ represents the exposure time needed to satisfy the specified SNR requirement. $T_\mathrm{slewing}$ denotes the time required for the telescope to adjust its orientation and stabilize its position when transitioning to the next sky region and ceil is the ceiling operator.\\

\begin{figure} [ht]
\begin{center}
\begin{tabular}{c} 
\includegraphics[height=7cm]{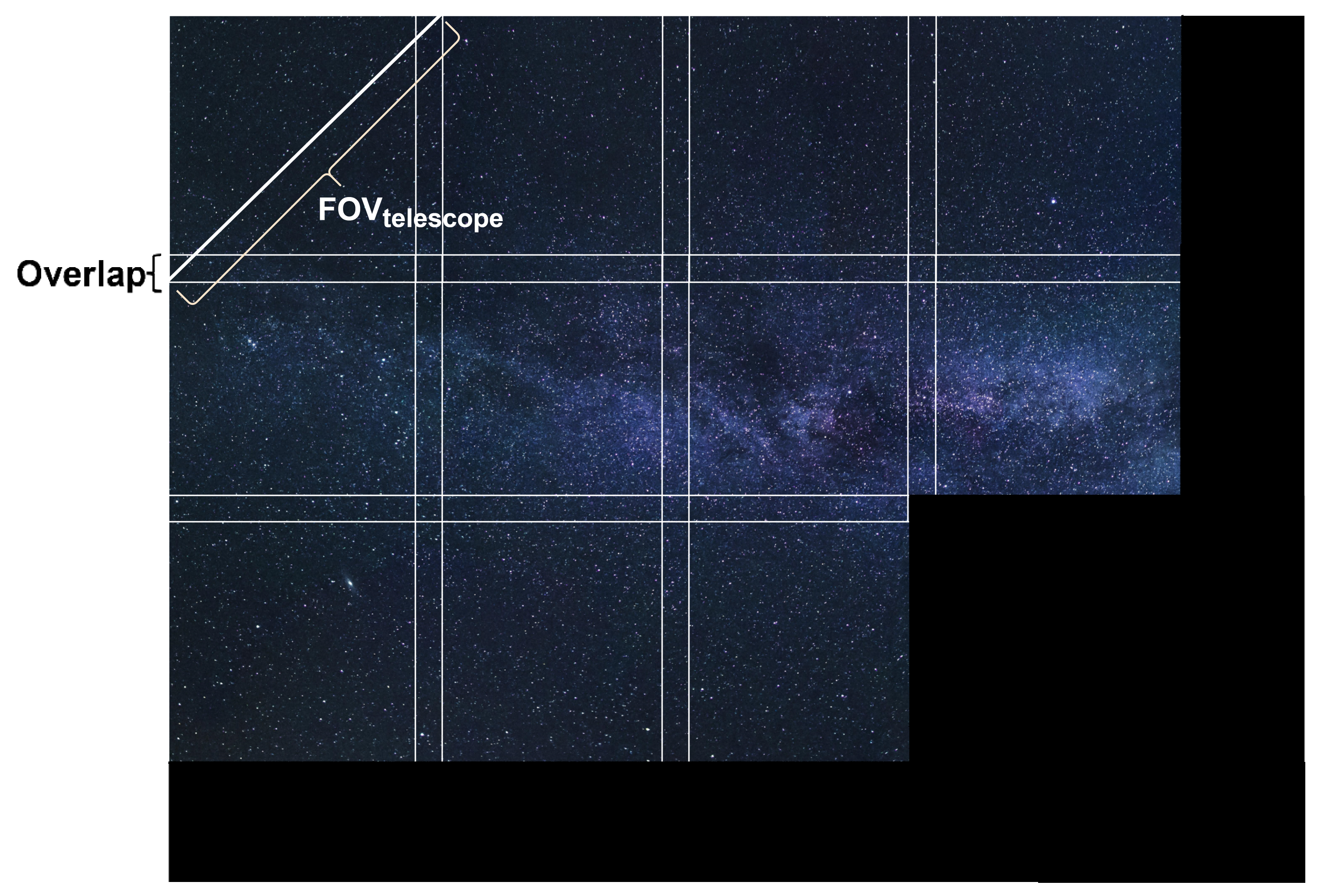}
\end{tabular}
\end{center}
\caption{\label{fig:5} 
The schematic diagram illustrates the overall field of view of a telescope array consisting of multiple telescopes. The region where the pixels overlap is referred to as the overlap area, defined as "overlap," which represents the shared pixel range of images obtained from adjacent positions in telescope observations. The maximum field of view of the telescope corresponds to the diagonal range of the image obtained from a single observation with an individual telescope.}
\end{figure}

It should be noted that since several pixels are in the overlap region to generate final observation images as shown in Figure~\ref{fig:5}, we define $ \Omega_\mathrm{overlap}$ to describe that problem, as defined in Equation~\ref{equation9}:
\begin{equation}
\begin{aligned}
\label{equation9}
\Omega_\mathrm{overlap} = (m\cdot (n-1)+n\cdot (m-1)) \cdot Fov_\mathrm{overlap\ width} \cdot  \frac{Fov_\mathrm{telescope}}{\sqrt{2}} \\ -3\cdot (m-1) \cdot (n-1) \cdot Fov_\mathrm{overlap\ width}^{2},
\end{aligned}
\end{equation}
$\Omega_\mathrm{overlap}$ represents the solid angle of field of view affected by overlapping observed sky regions, where the total number of observed sky areas obtained by the telescope array's optical system is predetermined as $m$ rows and $n$ columns, $Fov_\mathrm{overlap\ width}$ denotes the width of the overlapping area within the field of view, corresponding to the "Overlap" shown in Figure~\ref{fig:5}, while $Fov_\mathrm{telescope}$ represents the full field of view of the telescope. The overlapping field of view resulting from the pointing of different sky regions provides important information for the post-processing of images, particularly in the calibration and validation of observational data. Furthermore, this overlap plays a significant role when there is a desire to conduct in-depth observations of specific areas of the sky, facilitating the detection of transient astronomical events. By combining multiple observations, the overlapping field not only improves overall data quality but also effectively reduces noise and enhances the signal, thereby providing a more reliable foundation for scientific research.\\
 
After obtaining the overlapping field of view, we can obtain the total solid angle field of view $\Omega_\mathrm{all fov}$ with Equation~\ref{equation10}:
\begin{equation}
\begin{aligned}
\label{equation10}
\Omega_\mathrm{all fov} = {(\frac{Fov_\mathrm{telescope}}{\sqrt{2}})} ^{2} \times N_\mathrm{pointing}- \Omega_\mathrm{overlap},
\end{aligned}
\end{equation}
$N_\mathrm{pointing}$ is the total number of pointings during observation and also the total number of stitched images. We iterate $N_\mathrm{pointing}$ to ensure that the observational field of view of the telescope array can meet the scientific observation requirements. In summary, by using Equations~\ref{stack} to Equation~\ref{equation10}, we determine the actual total number of telescopes $N_\mathrm{telescope}$ required to satisfy the scientific observation needs.\\

It becomes challenging to accurately quantify the cost of the telescope and camera, due to various factors such as the system design or construction technology. In our framework, we establish a benchmark using several off-the-shelf telescopes with a diameter of 1 meter. The aperture size of a telescope influences its cost by the k index ratio. It is important to note that the proportion of manufacturing costs may differ for various telescope configurations. However, for the purpose of this paper, we utilize Equation~\ref{equation11} and \ref{equation12} to calculate the cost for each telescope as well as the entire telescope array:
\begin{equation}
\begin{aligned}
\label{equation11}
C_\mathrm{telescope} = C_\mathrm{telescope, 0} \cdot  D^{k},
\end{aligned}
\end{equation}
\begin{equation}
\begin{aligned}
\label{equation12}
C_\mathrm{total} = (C_\mathrm{telescope}+C_\mathrm{maintain}+C_\mathrm{camera})\cdot N_\mathrm{telescope},
\end{aligned}
\end{equation}

where $C_\mathrm{telescope}$ represents the cost of a single telescope with an aperture diameter of D, while $C_\mathrm{telescope, 0}$ denotes the cost of a single telescope with an aperture diameter of 1 meter. $D$ represents the ratio of the aperture size to the aperture size of the prototype telescope (1 m). $k$ is the exponential transformation parameter that determines the cost, which will increase as the aperture of this type of telescope system increases. The total cost $C_\mathrm{total}$ of the optical system for a telescope array is determined by the telescope cost $C_\mathrm{telescope}$, the maintenance cost $C_\mathrm{maintain}$, the camera cost $C_\mathrm{camera}$, and the number of telescopes $N_\mathrm{telescope}$ in the array. In our framework, we assume a value of $k=2$ since we could reduce cost with mass production to construct small telescopes. An additional $15\%$ of the construction cost is allocated for annual maintenance expenses. The cost of the camera is directly proportional to the size of the detector, which is determined by the dimensions of the image plane. Image plane influenced by parameters such as the aperture, F-number, and field of view of the telescope's optical system. Given these parameters, we can calculate the size of the imaging plane  of the current system using ZEMAX, which corresponds to the size of the camera. By combining this information with the approximate production cost per unit of the camera detector, we can derive the production cost of the camera corresponding to the telescope's optical system. The price of the camera can be found in Table \ref{tab:2} and we adjust the price of the camera according to the detector size. Incorporating the aforementioned parameters, we could conduct a comprehensive search across the entire parameter space to evaluate various configurations of telescopes and cameras. By comparing the total costs, we identify the most cost-effective parameter set, ensuring that the observational capabilities of the telescope array meet the actual scientific observation requirements.\\

\subsubsection{Fast Performance Evaluation with the Back Propagation Neural Network}
\label{sec242}
In the brute-force search algorithm, the optimizer continuously adjusts the telescope parameters by interacting with the optical system simulator and the exposure time calculator to obtain corresponding simulated images and compute the SNR. However, the process of dynamically invoking ZEMAX to compute PSFs at various imaging positions and returning the results is time-consuming. Moreover, even minor adjustments to environmental parameters necessitate recalculation of the results, and the algorithm lacks the ability to generalize. If the SNR obtained under different parameters through brute-force search are directly stored as a multi-dimensional array, subsequent calls would require searching within this array, which will face challenges when new sets of parameters are required for performance evaluation.\\

Considering that neural networks can effectively model complex functions and possess certain generalization capabilities, and recognizing that the SNR of the simulated images is influenced by the telescope's optical system parameters as well as the observational environment parameters, we propose using a BP neural network to model this relationship. By using the neural network algorithm trained with data obtained by brute force search, the corresponding SNR can be quickly calculated within the range of sampling parameters, significantly speeding up the retrieval speed of the algorithm within the parameter space. Moreover, by simply inputting the relevant telescope optical system parameters and observational environment parameters, the desired results of SNR can be conveniently obtained, simplifying repeated data computation and providing generalization capabilities that direct data storage does not possess.\\

\begin{figure*}[ht]
\includegraphics[height=2.4cm]{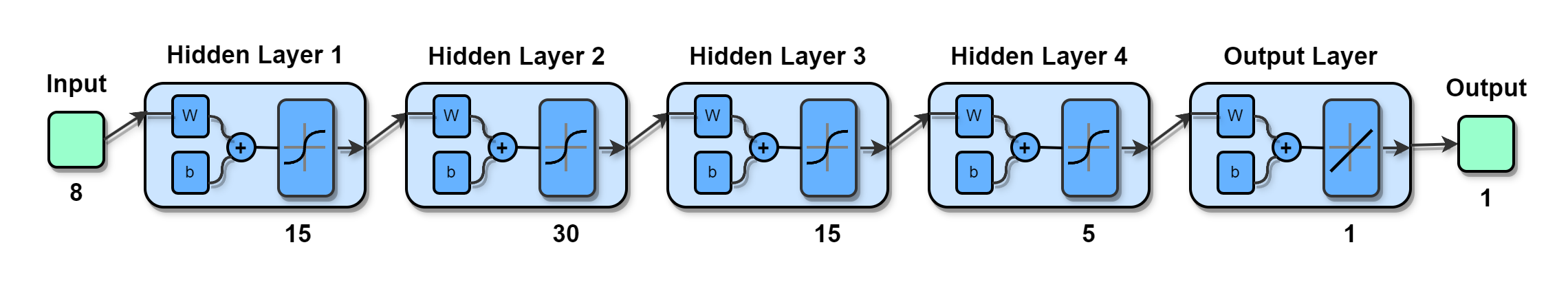}
\caption{\label{fig:6} The structure of the BP neural network used for performance estimation according to design parameters in this paper. The input layer comprises 8 neurons, which are followed by four hidden layers with neuron counts of 15, 30, 15, and 5, respectively, culminating in a single output result. The connections between the neurons in each layer are parameterized by the corresponding weight matrices (W) and bias vectors (b), which are adjusted during the network's training process.}
\end{figure*}

In this paper, we use the BP neural network for performance evaluation. The BP neural network is a well-established artificial neural network model comprising three main components: an input layer, hidden layers, and an output layer. The input layer receives initial data, while the hidden layers perform weighted calculations and apply activation functions to this data. Finally, the output layer produces the computed result. In this study, we utilize the BP neural network to establish a functional relationship between the parameters and the SNR of celestial objects. Specifically, our BP neural network takes eight parameters describing the observation conditions as inputs and outputs the SNR of celestial objects on the simulated images. We have built the BP neural network using MATLAB. The architecture of the neural network, as illustrated in Figure~\ref{fig:6}, consists of five hidden layers \citep{zhong2012application, wang2020bp}.\\

\begin{figure} [ht]
\begin{center}
\begin{tabular}{c} 
\includegraphics[height=10cm]{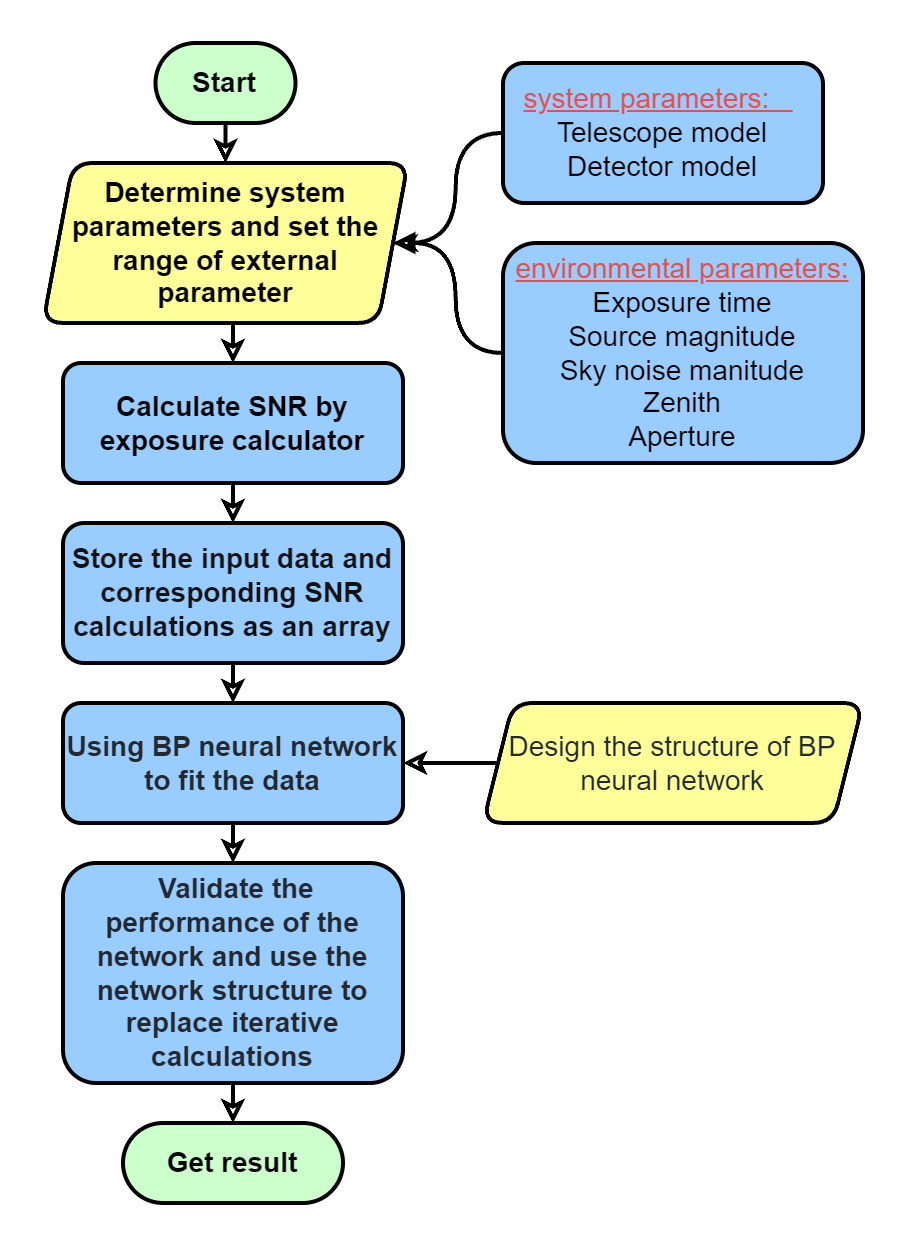}
\end{tabular}
\end{center}
\caption{ \label{fig:7} 
The procedure for acquiring data to train the BP neural network involves a systematic approach. We can conduct multiple iterations to collect the necessary training data, adjusting the parameters and expanding the dataset as required.}
\end{figure}

Data collection for training the BP neural network is conducted within specific parameter ranges relevant to a given task, as outlined in Figure~\ref{fig:7}. The training process begins with the use of an optical system simulator and an exposure time calculator. These tools utilize sampled parameter sets to calculate the SNR of celestial objects, creating the initial dataset for training the BP neural network. A primary challenge of using the BP neural network as a rapid data retrieval tool is the need for pre-collection of training data through interactive computations, which can be time-intensive. Nonetheless, once the network is trained, it can substitute the interactive process of SNR computation. This traditionally involves employing software like ZEMAX to determine the system's PSF and MATLAB for calculating SNR of simulated images. The BP neural network's ability to generalize enables it to assess the optical system's performance across the input environment's parameter space range. Although extracting SNR data from a multidimensional array is somewhat more intricate than performing direct interactive computations, the BP neural network provides substantial enhancements in processing speed. Section \ref{sec3} will provide a detailed comparison of retrieval efficiency in a specific case.\\

\section{Parameter Optimization For The SiTian Project}
\label{sec3} 

In this section, we demonstrate the application of our framework for parameter optimization in the SiTian Project. This ambitious ground-based telescope array initiative comprises dozens of small sky survey telescopes distributed across China and other parts of the world, complemented by at least three 4-meter telescopes dedicated to follow-up spectroscopy. The SiTian Project aims to observe a diverse range of celestial objects, including Active Galactic Nuclei, quasars, variable stars, planets, asteroids, and microlensing events. To achieve its scientific objectives, SiTian needs to scan a minimum of 10,000 square degrees of the sky every 30 minutes. The project employs simultaneous observation in three optical wavebands, with each unit consisting of three telescopes. SiTian's detection limit is set at 21 mag, allowing for the observation of relatively faint celestial objects. Since there are many free parameters need to be defined in the SiTian Project, we plan to use our framework to obtain parameters of the SiTian Project when the observation waveband is the V band.\\

Eight prototype telescopes are designed for the SiTian Project. These telescopes encompass a Schmidt telescope with F number of 2, Primary Focus telescopes (PFC) with F number of 1.23, 1.76, 2.8, and 2, Ritchey-Chrétien (RC) telescopes with F number of 3 and 4, Three Mirror Anastigmatic (TMA) telescope with F numbe of 3 as shown in Table~\ref{tab:telescopes}. The structure and materials of the telescopes could lead to different cost characteristics, which need to be assessed and adjusted by optical design experts. In this study, we do not address these details explicitly. The differences in PSF of telescopes and image quality can affect the exposure time needed to achieve scientific observation goals, thereby influencing the cost. Besides, we assign a relative cost to each telescope with a 1-meter aperture based on its field of view, computed by linear interpolation, with cost values ranging from $6 \times 10^{6}$ to $1.2 \times 10^{7}$ corresponding to field of view ranging from 3 degrees to 7 degrees. It is important to note that the specific cost values can be adjusted to reflect real conditions, and the numerical values provided in this paper are for reference purposes only. \\

\begin{table}[htbp]
\caption{Parameters of telescopes with different design used in this paper.} 
\label{tab:telescopes}
\begin{tabular}{@{} p{1.1cm}lll p{2.5cm} @{}}
\toprule
Serial    & Telescope model  & F\#  & FOV (deg) & Baseline cost of 1-meter(Relative) \\
\midrule
1      & Schmidt telescope            & 2    & 7      & $1.2 \times 10^{7}$                                                                  \\ 
2      & Primary Focus telescope      & 1.23 & 4.8    & $8.7 \times 10^{6}$                                                                   \\
3      & Primary Focus telescope      & 1.76 & 4.2    & $7.8 \times 10^{6}$                                                                   \\ 
4      & Primary Focus telescope      & 2.8  & 2      & $4.5 \times 10^{6}$                                                                   \\ 
5      & Primary Focus telescope      & 2    & 6      & $1.05 \times 10^{7}$                                                                  \\
6      & Ritchey-Chrétien telescope   & 3    & 4      & $7.5 \times 10^{6}$                                                                   \\ 
7      & Ritchey-Chrétien telescope   & 4    & 3      & $6 \times 10^{6}$                                                                   \\ 
8      & Three Mirror Anastigmatic telescope & 3    & 4  & $7.5 \times 10^{6}$                                                                   \\ 
\botrule
Cost setting & \multicolumn{4}{l}{\begin{tabular}[c]{@{}l@{}}The cost changes with the aperture with the k index.\\ In this paper, k=2, see Equations~\ref{equation11}\end{tabular}}                             
\\ \botrule
\end{tabular}

\end{table}

Using the defined hardware and environmental parameters, we have initially constructed a coarse grid of various parameters and have employed a brute force search method to evaluate the system's performance in the coarse grid. The parameter ranges and grids defined by our approach are detailed in Table~\ref{tab:set of the BP}. As shown in the table, during the brute force search process, the aperture interval for each telescope was set to 0.1 m, and the sampling interval for exposure time has been set to 2 seconds. Collecting all results for 8 different telescope prototypes under all environmental conditions has taken a total of 20.1 hours.\\

\begin{table}[ht]
\caption{Parameters, their respective ranges and number of sampling points in the training set of the BP neural network.} 
\label{tab:set of the BP}
\begin{tabular}{@{} llll p{2.5cm} @{}}
\toprule
Input parameters         & Range         & Sampling number  \\ 
\midrule
Seeing FWHM (")          & 1-2           & 6    \\
Zenith angle (°)         & 25-35         & 6       \\ 
Exposure time (min)      & 0-10          & 301        \\
Source magnitude (mag)        & 21        & 1          \\ 
Sky back magnitude (mag/rad)  & 22.3      & 1         \\ 
Aperture size  (m)       & 0.5–1.5        & 11    \\ 
Fov-x  (°)               & 0              & 1           \\ 
Fov-y    (°)     & 0-half field of view   & 6      \\ 
\botrule
\end{tabular}
\end{table}

\begin{figure}[htbp]
    \centering
    \begin{minipage}[t]{0.41\textwidth}
        \centering
        \includegraphics[width=\textwidth]{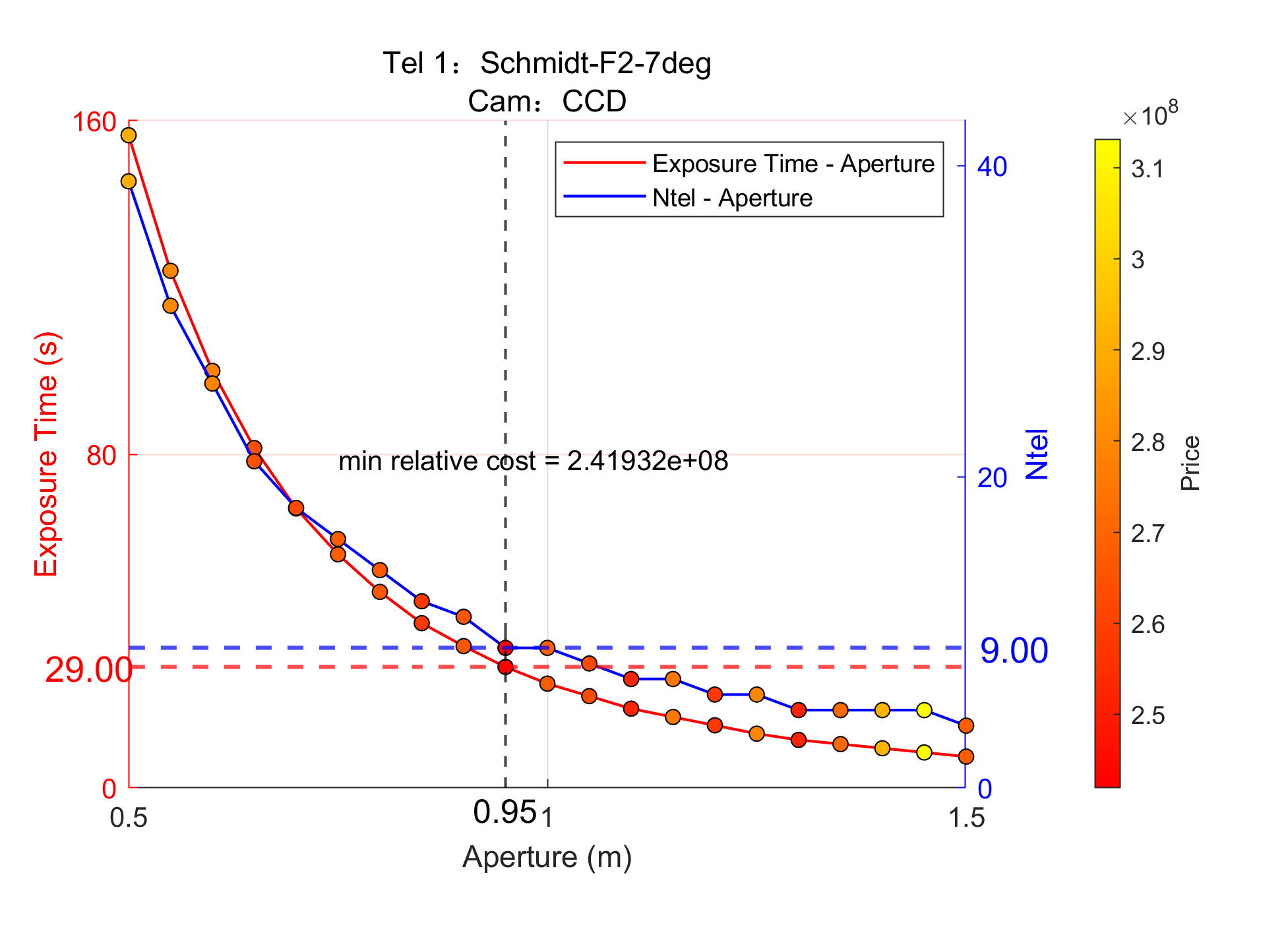}
    \end{minipage}
    \begin{minipage}[t]{0.41\textwidth}
        \centering
        \includegraphics[width=\textwidth]{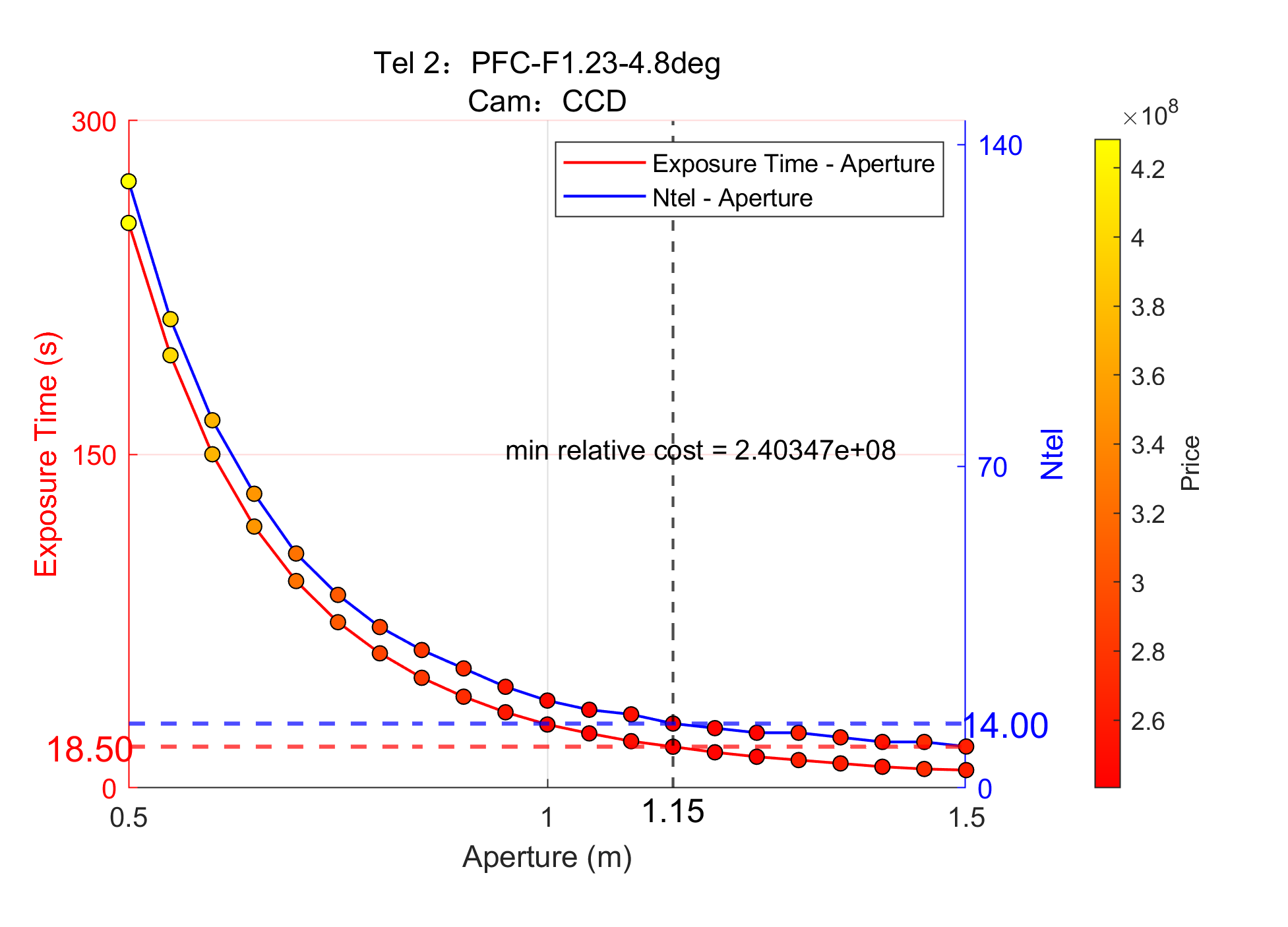}
    \end{minipage}
    
    \begin{minipage}[t]{0.41\textwidth}
        \centering
        \includegraphics[width=\textwidth]{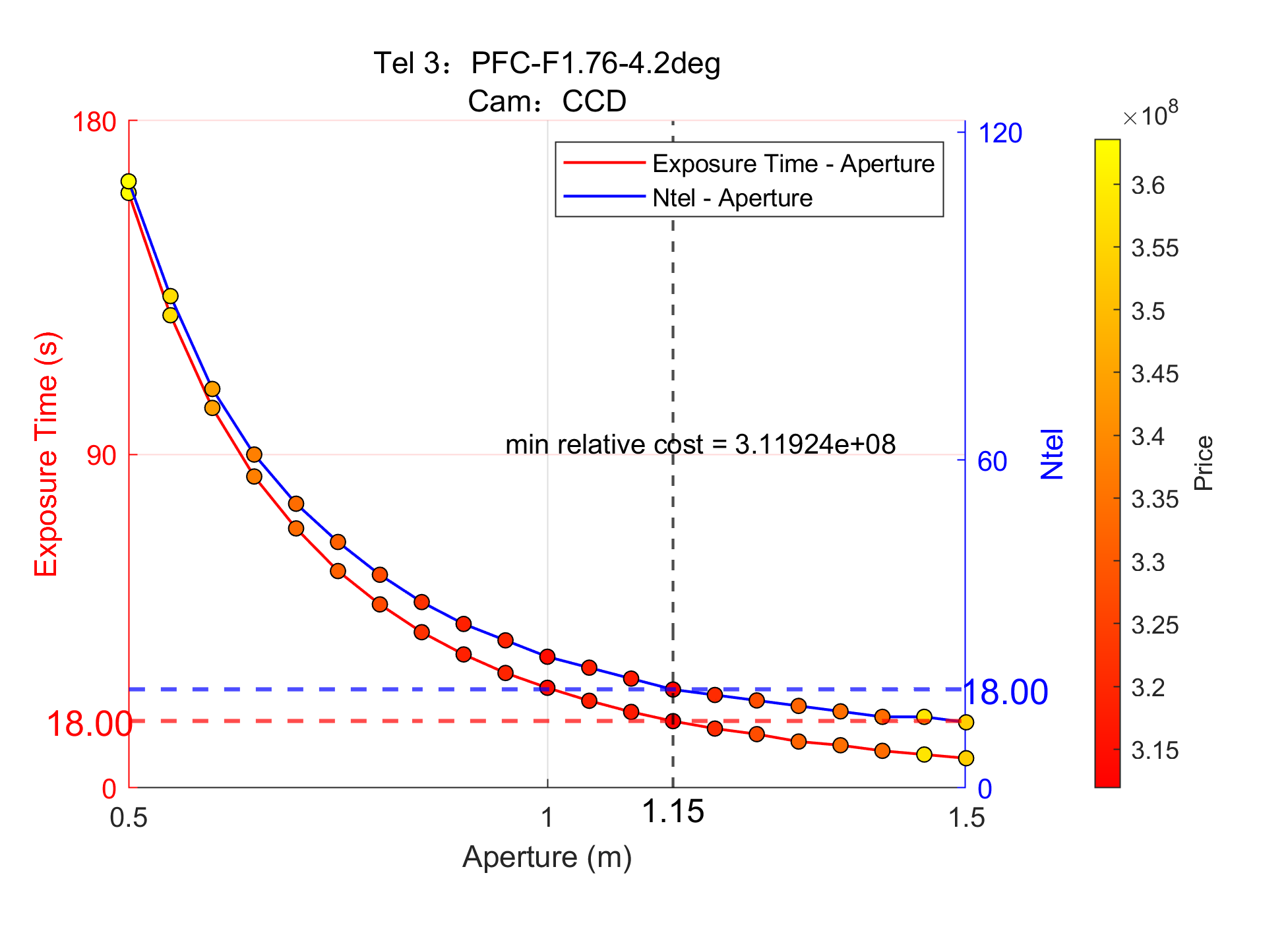}
    \end{minipage}
    \begin{minipage}[t]{0.41\textwidth}
        \centering
        \includegraphics[width=\textwidth]{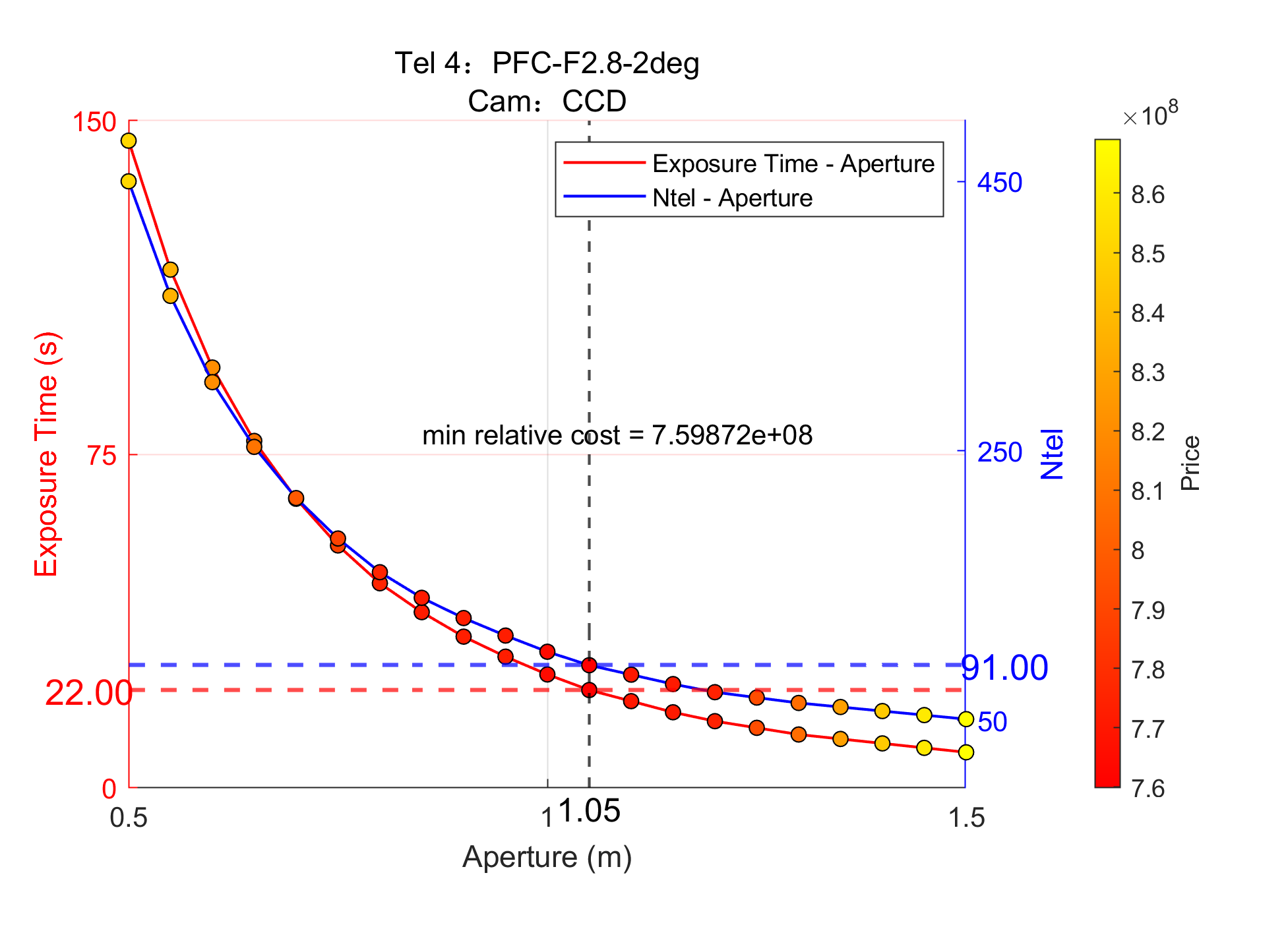}
    \end{minipage}
    
    \begin{minipage}[t]{0.41\textwidth}
        \centering
        \includegraphics[width=\textwidth]{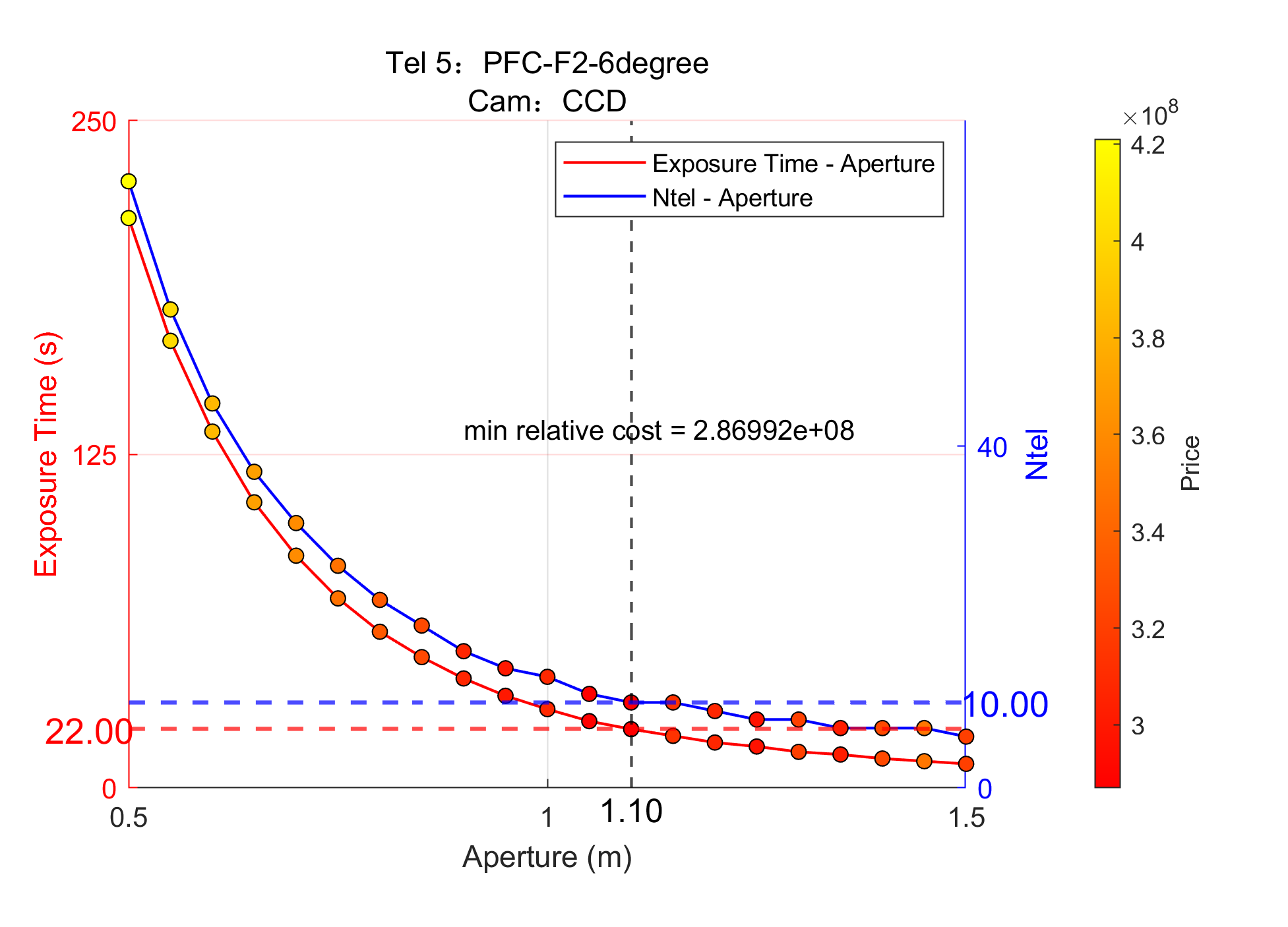}
    \end{minipage}
    \begin{minipage}[t]{0.41\textwidth}
        \centering
        \includegraphics[width=\textwidth]{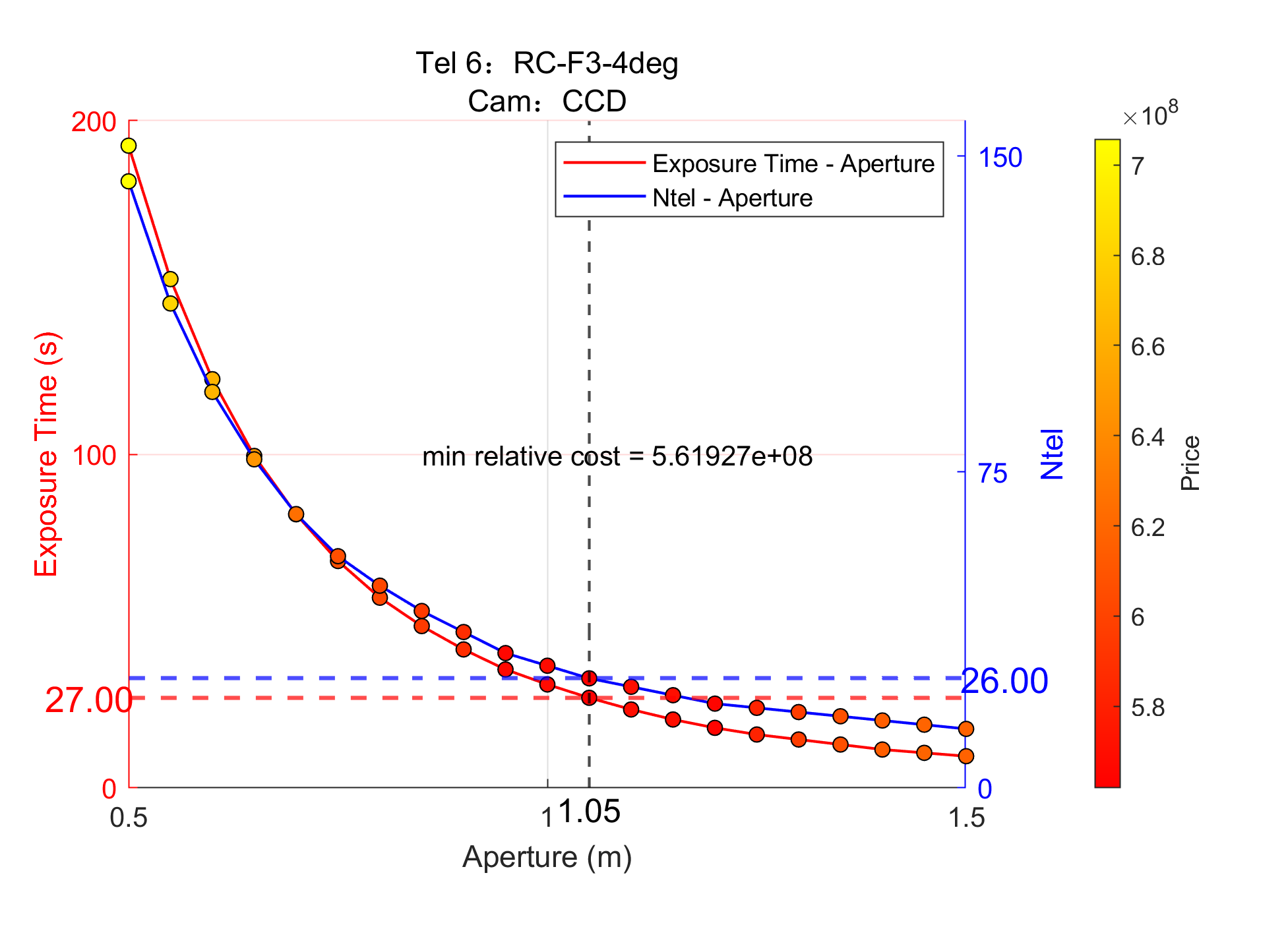}
    \end{minipage}
    
    \centering
    \begin{minipage}[t]{0.41\textwidth}
        \centering
        \includegraphics[width=\textwidth]{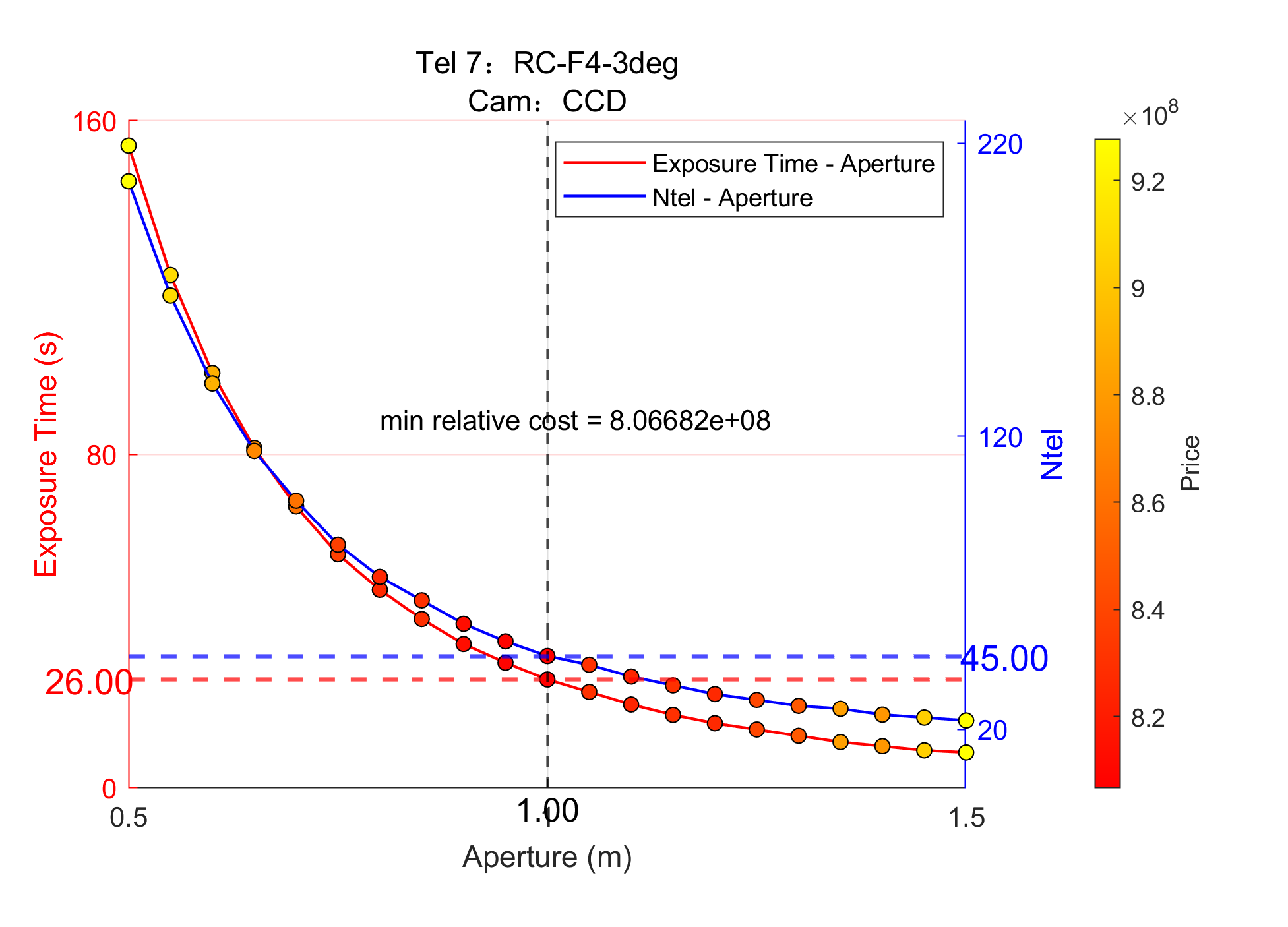}
    \end{minipage}
    \begin{minipage}[t]{0.41\textwidth}
        \centering
        \includegraphics[width=\textwidth]{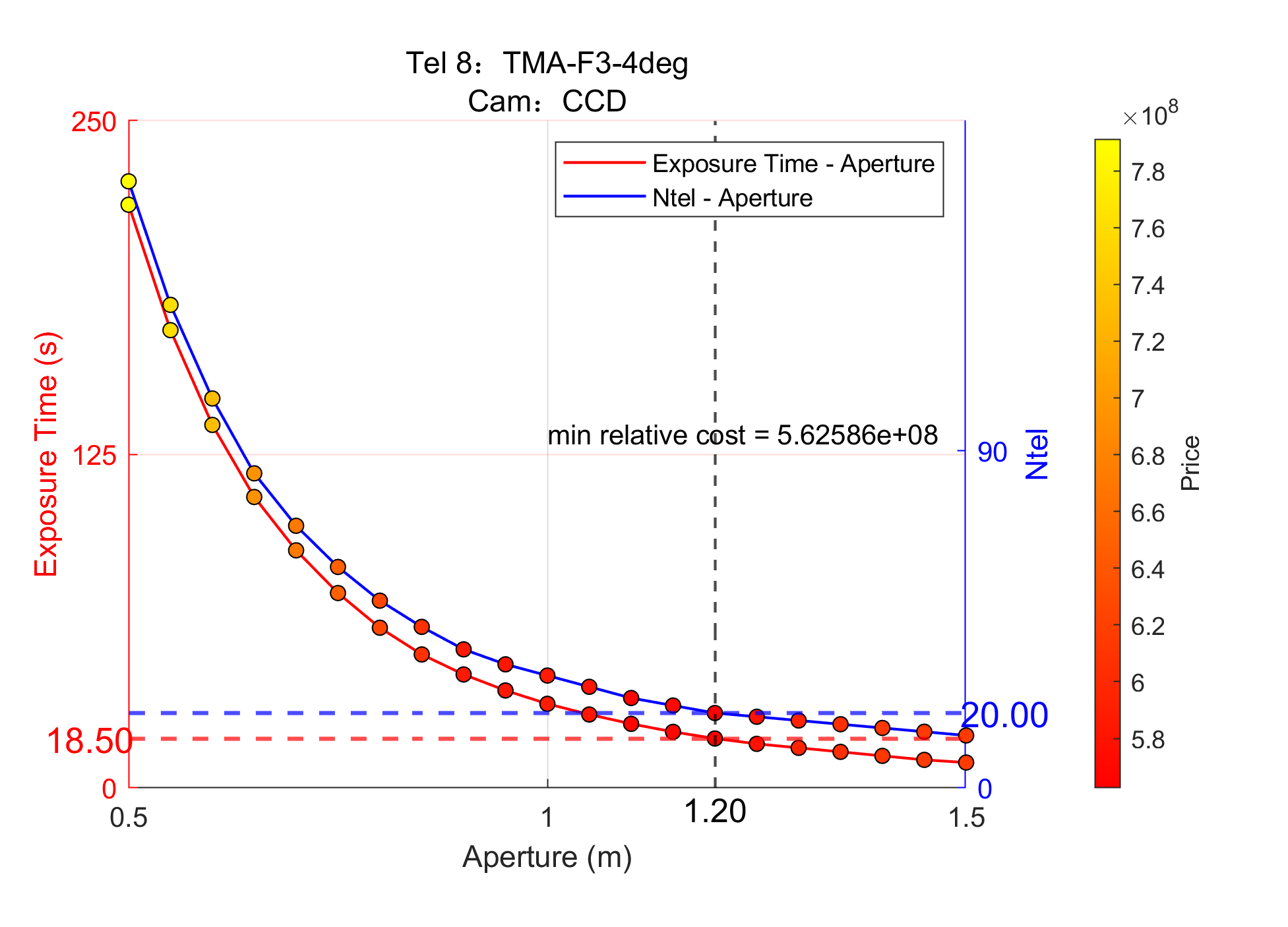}
    \end{minipage}
    \caption{The figure shows the results of the SiTian telescope array project's search in the V band using a BP neural network algorithm. The red and blue configuration curves illustrate the relationship between exposure time and the number and aperture of telescope units. This allows us to determine the minimum exposure time required for each telescope model to meet the observation area and SNR requirements at different apertures. Additionally, we have calculated the number of individual units needed in the array system to derive the configuration parameters that meet scientific requirements. The coordinates on the left and right sides of each subplot represent exposure time (s) and the number of telescopes (Ntel), respectively, while the horizontal axis denotes the telescope aperture (m), ranging from 0.5 m to 1.5 m with a sampling interval of 0.05 m and an exposure time interval of 0.5 seconds. The dashed lines corresponding to each axis indicate the configuration information with the minimum relative cost.}
    \label{fig:10}
\end{figure}

With results obtained by interactive brute force search method, we have trained a BP neural network for each of the 8 telescope prototypes, taking a total of 2.2 hours. As described in Section ~\ref{sec242}, with eight input parameters listed in Table~\ref{tab:set of the BP}, the trained neural network can replace the part of the algorithm that interacts with ZEMAX to calculate the simulated image's SNR. Utilizing the generalization capability of the BP neural network, we have further investigated the cost of SiTian under different configurations. The aperture interval for each telescope has been set as 0.05 m, and the sampling interval for exposure time has been set as 0.5 seconds, making the retrieval interval more refined. In Figure \ref{fig:10}, we present the retrieval results for 8 different configurations under the following environmental parameters: seeing of 1.5", zenith angle of 30°, celestial magnitude of 21, and sky background magnitude of 22.3. For each aperture, we can calculate the minimum exposure time required to meet the scientific observation needs of the SiTian project (which involves surveying 10,000 square degrees of the sky within 30 minutes while ensuring an imaging SNR of no less than 10) and determine the corresponding number of telescope units needed. Using the node parameters associated with each aperture, we can compute the most cost-effective configuration for specific combinations of telescope types and cameras, which are indicated by dashed lines marking their positions on the various parameter axes in the figure. By comparing the lowest costs of all configurations, we can preliminarily identify the most cost-effective optical system configuration for the telescope array under current conditions. Corresponding to the telescope models listed in Table \ref{tab:telescopes}, The lowest relative costs are $2.41932 \times 10^8$, $2.40347 \times 10^8$, $3.11924 \times 10^8$, $7.59872 \times 10^8$, $2.86992 \times 10^8$, $5.61927 \times 10^8$, $8.06682 \times 10^8$, and $5.62586 \times 10^8$. In this figure, the retrieval results based on the BP neural network show that selecting a PFC-F1.23-4.8deg telescope with an aperture of 1.15 m as the optical system in the array, and employing a survey strategy with 14 telescopes and a single exposure time of 18.5 seconds, is optimal.\\

\begin{figure} [ht]
\begin{center}
\begin{tabular}{c} 
\includegraphics[height=7.4cm]{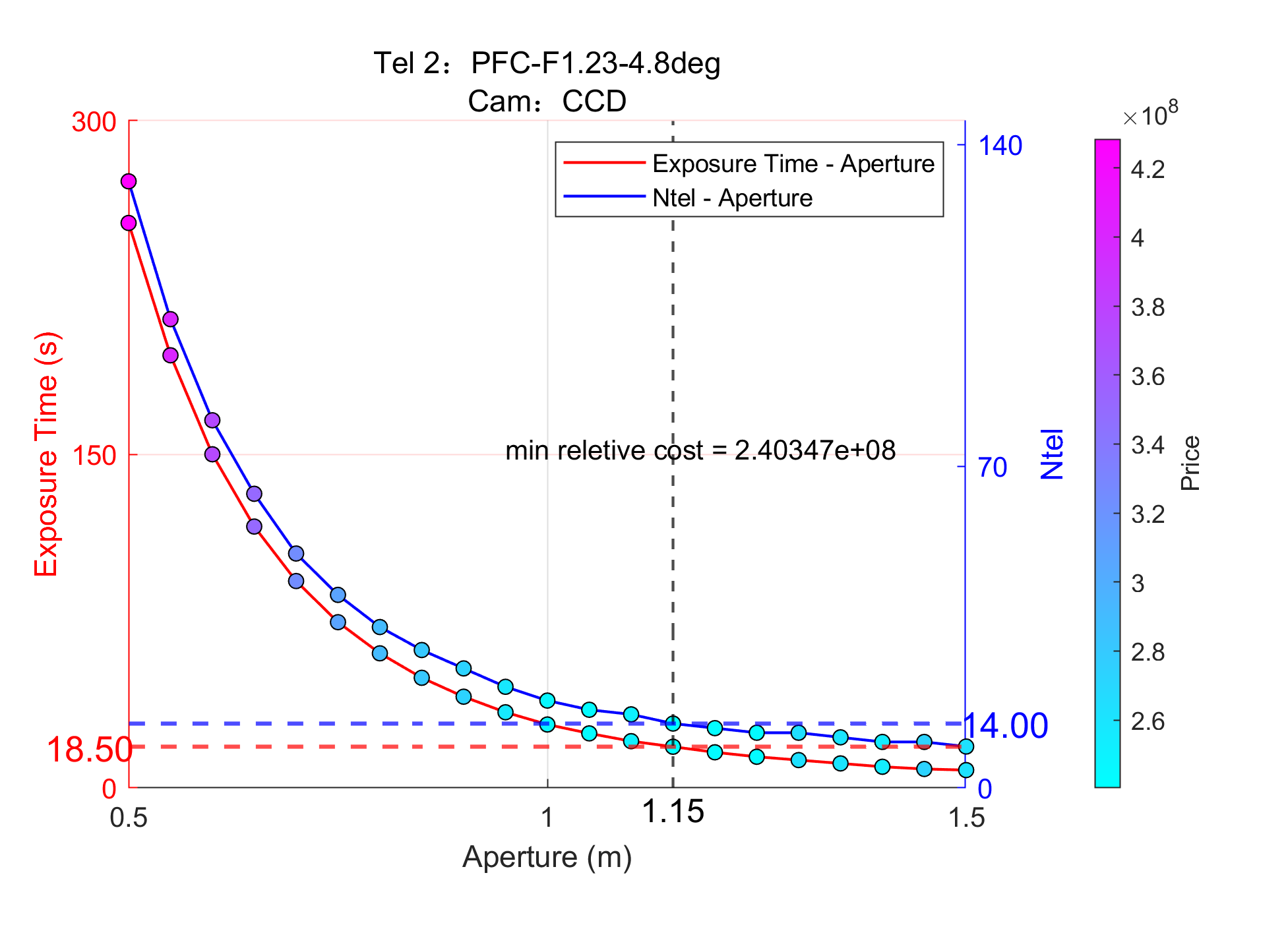}
\end{tabular}
\end{center}
\caption{ \label{fig:force search result} 
The telescope array configuration retrieval results for the PFC-F1.23-4.8deg telescope were obtained using an interactive brute-force search method. The configuration information of all nodes and the minimum cost obtained through computation are completely consistent with the results obtained with the BP neural network.}
\end{figure}

To verify the accuracy of the BP neural network's computational results, we conducted a brute force search for the PFC-F1.23-4.8deg telescope under the same environmental parameters and sampling grid. As shown in Figure \ref{fig:force search result}, it can be observed that the search results of the BP neural network are consistent with those obtained by the interactive brute force search method. However, with a finer retrieval grid for aperture and exposure time, using the brute force search method to retrieve all environmental conditions and telescope prototypes listed in Table~\ref{tab:set of the BP} would take approximately 25.4 hours, with each evaluation of a new configuration, similar to a subplot in Figure \ref{fig:10}, taking about 318 seconds, and the results lack generalization. In contrast, using the BP neural network to evaluate a new configuration within the sampling range takes only about 14 seconds to obtain results.\\

\begin{figure}[htbp]
    \centering
    \begin{minipage}[t]{0.48\textwidth}
        \centering
        \includegraphics[width=\textwidth]{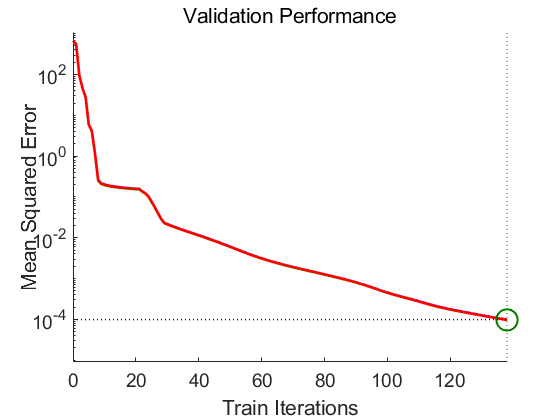}
        \textbf{(a)}
        \label{fig:a}
    \end{minipage}
    \hfill
    \begin{minipage}[t]{0.48\textwidth}
        \centering
        \includegraphics[width=\textwidth]{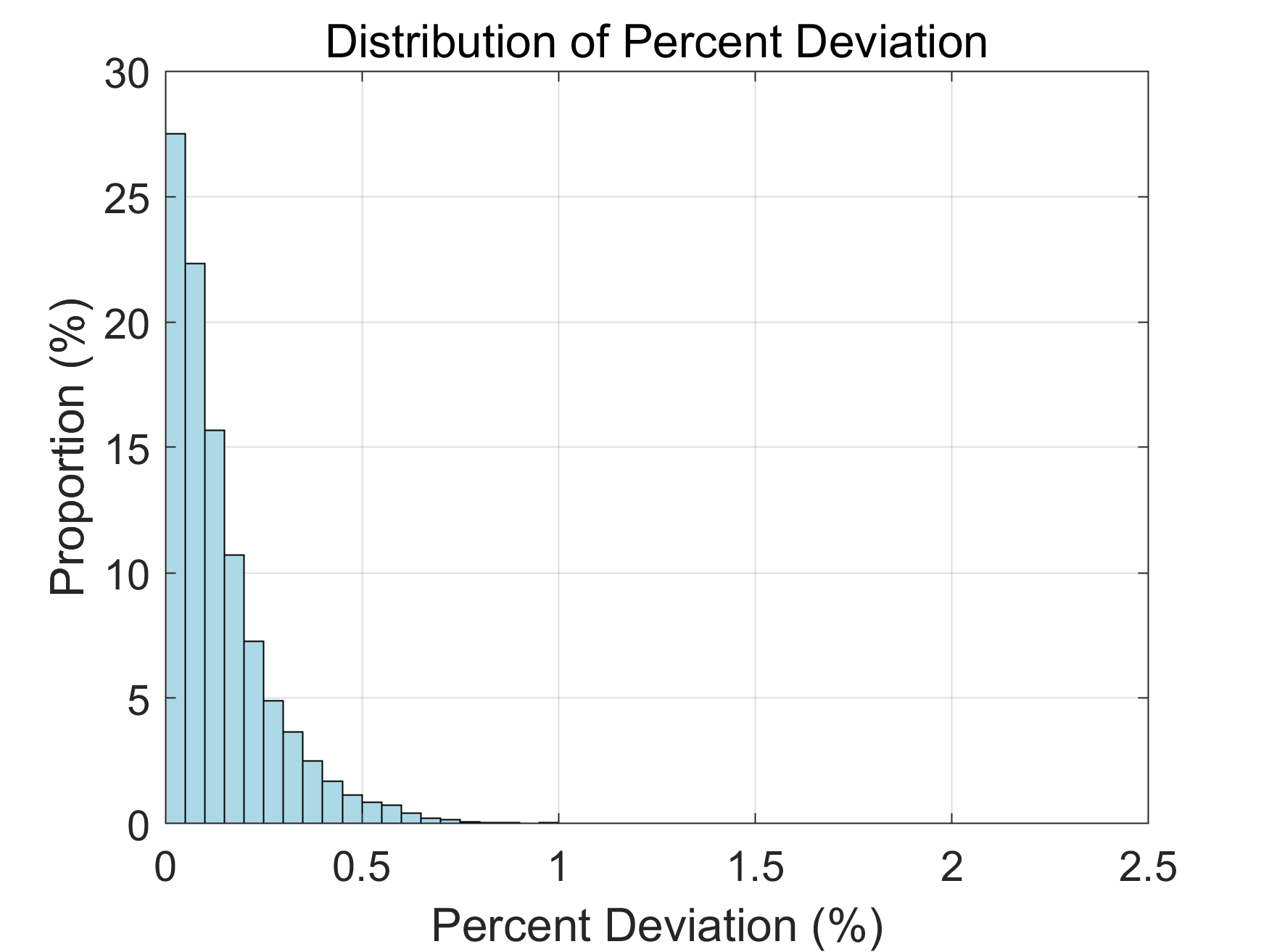}
        \textbf{(b)}
        \label{fig:b}
    \end{minipage}
    \caption{\label{fig:bp_train} When using the PFC-F1.23-4.8deg telescope as the optical system in the array, the training process of the BP neural network and the evaluated images are shown. (a) is an image of the neural network training process, with the horizontal axis representing the number of iterations in network training and the vertical axis representing the mean squared error (MSE) of the network data fitting. The network converges after 138 iterations, with the MSE reaching the threshold of $1 \times 10^{-4}$. (b) is a bar chart comparing the SNR results obtained through the BP neural network from 100,000 sets of randomly sampled input data with the SNR results of simulated images obtained interactively via ZEMAX. The percentage deviation between the SNR values calculated by the BP neural network and those obtained by the exposure time calculator does not exceed 1.5\%, with 99.2\% of the data having an relative error percentage less than 1.0\%.}
\end{figure}

In Figure \ref{fig:bp_train}(a), we further present the training results of BP neural network when using the telescope prototype PFC-F1.23-4.8deg. We have trained the network until the mean square loss of the network reached our set loss threshold, which is $1 \times 10^{-4}$. To evaluate the fitting results, we randomly generated 100,000 new parameter sets within the specified range. The SNR of celestial objects is calculated using these parameter sets and compared with the SNR obtained from the optical system simulator and the exposure time calculator. The deviation results between the two are shown in Figure \ref{fig:bp_train} (b). The calculation results of the BP neural network are close to those of the optical system simulator and exposure time calculator, with the maximum percentage deviation not exceeding 1.5\% and 99.2\% of the comparative data having a relative error of less than 1.0\%. This confirms the reliability of the BP neural network algorithm in fitting the SNR calculations.\\

It should be noted that, in image simulation step with the brute force search method, the presence of random noise often introduces a certain level of uncertainty into the results. When using the PFC-F1.23-4.8deg telescope paired with a custom CCD camera (with environmental parameters: seeing of 1.5", a zenith angle of 30°, celestial magnitude of 21, and sky background magnitude of 22.3), the search results using the BP neural network and brute force search method are detailed in Table \ref{tab:5}. Although there are slight differences between the two, they also demonstrate the effectiveness of the BP algorithm. Additionally, the neural network structure offers advantages such as convenient storage and retrieval and strong generalization capabilities. The data above indicate that we can effectively use neural networks to replace parts of the interactive calculation method for image's SNR computation. This substitution can enhance the speed of retrieval within the sampled parameter interval.\\

\begin{table}[htbp]
\caption{Parameters of the telescope array calculated by interactive brute-force search and a BP neural network with PFC-F1.23-4.8deg telescope/custom CCD (9 micron/pixel) as effective design. For data where the results from the two methods differ, they are represented as (interactive calculation result / neural network substituted calculation result). The most cost-effective strategy parameters correspond to an aperture of 1.15 m.} 
\label{tab:5}
\begin{tabular}{l p{1.5cm} p{1.5cm} l p{2.0cm} l}
\toprule
Aperture (m) & Exposure Time (s) & Telescope Number & Relative Cost & Observation Area ($\deg^2$) & SNR (min) \\
\midrule

1.50        & 8.0    & 9      & $2.6286\times 10^{8}$ & $1.0007\times 10^{4}$ & 10.31 / 10.27    \\

1.45        & 8.5    & 10     & $2.7293\times 10^{8}$ & $1.0002\times 10^{4}$ & 10.06 / 10.02    \\ 

1.40        & 9.5    & 10     & $2.5443\times 10^{8}$ & $1.0010\times 10^{4}$ & 10.04 / 10.02    \\ 

1.35        & 11.0   & 11     & $2.6024\times 10^{8}$ & $1.0005\times 10^{4}$ & 10.17   \\ 

1.30        & 12.5   & 12    & $2.6325\times 10^{8}$ & $1.0011\times 10^{4}$ & 10.17 / 10.18    \\

1.25        & 14.0   & 12    & $2.4339\times 10^{8}$ & $1.0006\times 10^{4}$ & 10.07 / 10.08   \\

1.20        & 16.0   & 13    & $2.4301\times 10^{8}$ & $1.0011\times 10^{4}$  & 10.06 / 10.07    \\ 

1.15        & 18.5   & 14    & $2.4035\times 10^{8}$ & $1.0005\times 10^{4}$  & 10.10 / 10.11   \\ 

1.10        & 21.0   & 16    & $2.5131\times 10^{8}$ & $1.0009\times 10^{4}$  & 10.02   \\ 

1.05        & 24.5   & 17    & $2.4329\times 10^{8}$ & $1.0001\times 10^{4}$   & 10.06 / 10.05   \\ 

1.00        & 28.5   & 19    & $2.4664\times 10^{8}$ & $1.0004\times 10^{4}$   & 10.01   \\ 

0.95        & 34.0   & 22    & $2.5774\times 10^{8}$ & $1.0006\times 10^{4}$   & 10.02 / 10.04   \\ 

0.90        & 41.0   & 26    & $2.7339\times 10^{8}$ & $1.0008\times 10^{4}$   & 10.03 / 10.01   \\

0.85        & 49.5   & 30    & $2.8136\times 10^{8}$ & $1.0007\times 10^{4}$  & 10.02 / 10.04   \\ 

0.80        & 60.5   & 35    & $2.9077\times 10^{8}$ & $1.0006\times 10^{4}$   & 10.03    \\ 

0.75        & 74.5   & 42    & $3.0668\times 10^{8}$ & $1.0002\times 10^{4}$   & 10.02 / 10.01   \\

0.70        & 93.0   & 51    & $3.2440\times 10^{8}$ & $1.0008\times 10^{4}$   & 10.02   \\

0.65        & 117.5  & 64    & $3.5099\times 10^{8}$ & $1.0000\times 10^{4}$  & 10.02 / 10.01    \\ 

0.60        & 150.0  & 80    & $3.7384\times 10^{8}$ & $1.0000\times 10^{4}$  & 10.00 / 10.01   \\

0.55        & 194.5  & 102   & $4.0052\times 10^{8}$ & $1.0008\times 10^{4}$ & 10.01 / 10.00    \\ 

0.50        & 254.0  & 132   & $4.2837\times 10^{8}$ & $1.0010\times 10^{4}$ & 10.00    \\ 
\botrule

\end{tabular}
\end{table}

\section{Conclusions And Future Works}
\label{sec4} 
In recent years, telescope array projects employing multiple wide-field telescopes for sky surveys have attracted considerable attention as they serve as effective observational tools for the study of time-domain astronomy. Designing an optimized telescope array presents significant challenges, owing to the multitude of parameters that need to be considered. This paper introduces a framework that automatically determines the optimal parameter set for a telescope array, thereby assisting optical experts in evaluating the capabilities of such optical system arrays. The framework is designed to assess the performance of the telescope array across varied parameter sets, with the goal of minimizing costs while satisfying specific scientific requirements and considering available instrumentation options. It consists of three key components: an exposure time calculator, an optical system simulator, and a system optimizer. The exposure time calculator inputs design parameters and test data to assess the total cost of a telescope array by evaluating the exposure time and the SNR of the system. Meanwhile, the optical system simulator produces PSFs for a given telescope based on its design specifications. Finally, the system optimizer probes the parameter space of the telescope array to identify the most efficient configurations.\\

We have evaluated our framework utilizing parameters from the SiTian project, and the findings demonstrate that it can effectively determine the most cost-efficient parameter configuration for a telescope array. This capability offers optical experts valuable assistance and serves as a useful reference. We have incorporated a BP neural network algorithm to model results derived from a brute force search, significantly enhancing the retrieval speed within the sampled parameter range. However, the current framework for estimating costs in telescope optical systems is hindered by the absence of a dedicated internal database and necessary precision. Instead, it depends on the knowledge and expertise of optical engineering professionals for the final comprehensive evaluation and selection, based on the specific requirements of scientific observation missions. To improve its computational capabilities, more data are required on the actual manufacturing costs of telescopes with varied structural designs and the effects of optomechanical structures on imaging quality. Future iterations of the framework aim to optimize the detailed cost estimation algorithms further, thereby offering more professional support to optical engineers. This enhancement is intended to refine the decision-making process and bolster the framework's applicability in diverse observational and structural scenarios. Through ongoing development and integration of additional data, the framework is expected to provide a robust tool for cost analysis and system optimization in telescope design.\\

\bibliography{sn-bibliography}

\end{document}